\documentclass[pss]{wiley2sp} 
\usepackage{amsmath}
\usepackage{bm}              
\usepackage{w-greek}         

\begin{document}

\title{Novel Multifunctional Materials Based on Oxide Thin Films and Artificial Heteroepitaxial Multilayers}

\titlerunning{Novel Multifunctional Materials Based on Oxide Thin Films and Artificial Heteroepitaxial Multilayers}

\author{%
  Matthias Opel\textsuperscript{\Ast,\textsf{\bfseries 1}},
  Stephan Gepr\"{a}gs\textsuperscript{\textsf{\bfseries 1}},
  Edwin P. Menzel\textsuperscript{\textsf{\bfseries 1,3}},
  Andrea Nielsen\textsuperscript{\textsf{\bfseries 1}},
  Daniel Reisinger\textsuperscript{\textsf{\bfseries 1}},
  Karl-Wilhelm Nielsen\textsuperscript{\textsf{\bfseries 1}},
  Andreas Brandlmaier\textsuperscript{\textsf{\bfseries 1}},
  Franz D. Czeschka\textsuperscript{\textsf{\bfseries 1}},
  Matthias Althammer\textsuperscript{\textsf{\bfseries 1}},
  Mathias Weiler\textsuperscript{\textsf{\bfseries 1}},
  Sebastian T.B. Goennenwein\textsuperscript{\textsf{\bfseries 1}},
  J\"{u}rgen Simon\textsuperscript{\textsf{\bfseries 2}},
  Matthias Svete\textsuperscript{\textsf{\bfseries 2}},
  Wentao Yu\textsuperscript{\textsf{\bfseries 2}},
  Sven-Martin H\"{u}hne\textsuperscript{\textsf{\bfseries 2}},
  Werner Mader\textsuperscript{\textsf{\bfseries 2}},
  Rudolf Gross\textsuperscript{\textsf{\bfseries 1,3}}}

\authorrunning{Matthias Opel et al.}

\mail{e-mail
  \textsf{opel@wmi.badw.de}, Phone: +49-89-289-14237, Fax: +49-89-289-14206}

\institute{%
  \textsuperscript{1}\,Walther-Mei{\ss}ner-Institut, Bayerische Akademie der Wissenschaften, 85748 Garching, Germany\\
  \textsuperscript{2}\,Institut f\"{u}r Anorganische Chemie, Universit\"{a}t Bonn, 53117 Bonn, Germany\\
  \textsuperscript{3}\,Physik-Department, Technische Universit\"{a}t M\"{u}nchen, 85748 Garching, Germany
}

\received{XXXX, revised XXXX, accepted XXXX} 
\published{XXXX} 


\abstract{%
\abstcol{%
Transition metal oxides show fascinating physical properties such as high
temperature superconductivity, ferro- and antiferromagnetism, ferroelectricity
or even multiferroicity. The enormous progress in oxide thin film technology
allows us to integrate these materials with semiconducting, normal conducting,
dielectric or non-linear optical oxides in complex oxide heterostructures,
providing the basis for novel multi-functional materials and various device
applications. Here, we report on the combination of ferromagnetic,
semiconducting, metallic, and dielectric materials properties in thin films and
artificial heterostructures using laser molecular beam epitaxy. We discuss
the fabrication and characterization of oxide-based ferromagnetic tunnel
junctions, transition metal-doped semiconductors, intrinsic multiferroics, and
artificial ferroelectric/ferromagetic heterostructures - the latter allow for
the detailed study of strain effects, forming the basis of spin-mechanics. For
characterization we use X-ray diffraction, SQUID magnetometry, magnetotransport
measurements, and advanced methods of transmission electron microscopy with the
goal to correlate macroscopic physical properties with the microstructure of
the thin films and heterostructures.}}

\titlefigure[height=6cm]{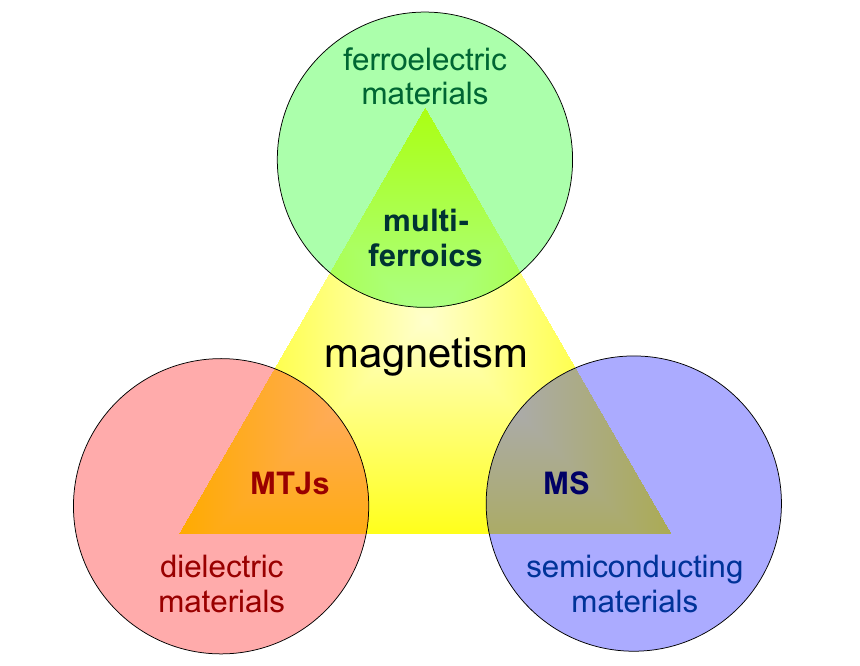}
\titlefigurecaption{
The combination of magnetic properties with dielectric, semiconducting, or
ferroelectric materials in one and the same material (e.g. magnetic
semiconductors (MS) or intrinsic multiferroics) as well as in artificial
heterostructures (e.g. ferromagnetic/dielectric heterostructures for magnetic
tunnel junctions (MTJs) or artificial multiferroic heterostructures) allows for
the design of materials with novel functionalities and provides the basis for
various device applications. }

\maketitle

\section{Introduction.}
 \label{sec:Intro}

Oxide materials are in the focus of intense research activities (see e.g.
Refs.~\cite{Blamire2009,Johnsson2008} for recent reviews) since more than two
decades. On the one hand, they show a large variety of interesting and in some
cases unique physical properties. Prominent examples are the cuprate
superconductors~\cite{Bednorz1986}, ferroelectrics~\cite{Jia1986}, non-linear
optical materials, electrically controllable dielectrics~\cite{Zimmermann2001},
the colossal magnetoresistance manganites~\cite{Helmolt1993,Jin1994}, or
multiferroic materials~\cite{Spaldin2005,Fiebig2005}. On the other hand, oxides
offer a broad potential for device applications. Although the underlying
physics of some oxide materials is still under debate (e.g. high-temperature
superconductivity), they are widely used in applications. A particularly
promising research direction is the integration of the outstanding properties
of different oxides in complex heterostructures to achieve materials with
improved or extended functionalities. Such materials may pave the way for novel
device structures. Notable examples are ferroelectric field-effect transistors,
controllable dielectric microwave devices, superconducting high-frequency
modules, magnetoresistive sensors, or electrically controllable and readable
magnetic random access memories.

An obvious approach for the realization of oxide materials with improved
functionality is the integration of two properties in one and the same
material. Well known examples are ferromagnetic
semiconductors~\cite{Ohno1996,Dietl2000,Venkatesan2004} combining
semiconducting and ferromagnetic properties or intrinsic
multiferroics~\cite{Spaldin2005,Fiebig2005} having a finite ferromagnetic and
ferroelectric order parameter. However, the realization of such intrinsically
multi-functional materials turns out to be difficult, in particular when room
temperature operation is required. Another promising approach for the
realization of multi-functional materials is the integration of oxides with
different properties in artificial heterostructures. This approach has become
feasible with the enormous progress in oxide thin film technology over the last
two decades. For example, today laser molecular beam epitaxy (laser-MBE)
employing \textit{in-situ} reflection high energy electron diffraction (RHEED)
allows us to grow oxide thin films with crystalline quality approaching
semiconductor standards~\cite{Gross2000,Gupta1990,Klein1999,Klein2000}.
Moreover, in close analogy to GaAs/AlAs heteroepitaxy it is possible to grow
complex heterostructures composed of different oxides on suitable substrates in
a layer-by-layer or block-by-block mode~\cite{Gross2000,Reisinger2003a}.

The successful heteroepitaxial growth of oxide materials in complex multilayers
opens various possibilities to deliberately tailor their physical properties
and to generate new functionalities. First, in analogy to semiconductor
heterostructures band gap engineering can be employed to tailor the band
structure at the interfaces. A prominent example is the system
LaAlO$_3$/SrTiO$_3$, forming a two-dimensional electron gas at the
interface~\cite{Ohtomo2004} which shows superconductivity below
200\,mK~\cite{Reyren2007}. Band bending effects due to interface charges may
also play an important role in ferromagnet/ferroelectric/ferromagnet tunnel
junctions, resulting in differences of the tunneling resistance dependent on
the polarization direction in the ferroelectric tunneling
barrier~\cite{Zhuravlev2005,Tsymbal2006}. Second, magnetic oxide
heterostructures may allow to tailor the magnetic interactions at the
interfaces thereby providing the basis for deliberate spin engineering. In
this way novel magnetic properties can be designed by varying the layer
structures. However, since the magnetic interactions in most cases are short
range, this approach requires good control of the interfaces on an atomic
scale, making controlled spin engineering very demanding. Third,
heteroepitaxial multilayers in most cases involve epitaxial coherency strain.
This can be used to deliberately tailor the materials properties in
heterostructures by strain. Strain effects have been successfully applied to
superconducting~\cite{Gupta1990,Gross1990},
magnetoresistive~\cite{Wiedenhorst1999,Wiedenhorst2000,Lu2000,Lu2005},
ferromagnetic~\cite{Goennenwein2008,Brandlmaier2008,Weiler2009}, and ferroelectric
materials~\cite{Choi2004,Bousquet2008}. Fourth, so-called asymmetric
superlattices such as BaTiO$_3$/SrTiO$_3$/CaTiO$_3$ can be used to break
inversion symmetry. In this way improved ferroelectric and non-linear optical
properties can be obtained~\cite{Sai2000,Lee2005,Warusawithana2003}. Finally,
superlattices of materials possessing different types of order parameter (e.g.
superconducting, ferromagnetic, ferroelectric) can be fabricated, allowing for
the coupling of the order parameters at the interfaces. Although this coupling
in many cases is not well understood, it led to applications in which a single
property, such as magnetization or electrical conductivity, can controllably be
turned on and off~\cite{Kanki2006,Salvador1999}. Along with many experimental
achievements, a growing number of theory predictions has been published
regarding artificial material systems based on heterostructures
\cite{Neaton2003,Pertsev2000,Okamoto2004,Junquera2003,Duan2006,Stengel2006}.

In this article, we restrict our discussion to heterostructures based on
magnetic oxides. We review our results obtained within the priority program
1157 of the German Research Foundation.
Magnetic materials and heterostructures have achieved broad
attention due to their interesting physics and their versatile application in
the field of magneto and spin electronics~\cite{Prinz1998,Wolf2001,Zutic2004}.
In this context ferromagnetic oxides are of particular interest since some of
them are expected to show an almost complete spin polarization at the Fermi
edge or are considered as ferromagnetic insulators. Interesting material
systems are magnetite (Fe$_3$O$_4$) and the double perovskites with composition
$A_2BB^\prime$O$_6$ ($A=$ alkaline earth, $B$, $B^\prime =$ magnetic or
nonmagnetic transition metal ion) as they show Curie temperatures well above
room temperature and band structure calculations predict half-metallic
behavior. This makes them well suited as electrode materials for spin injection
into semiconductors or the realization of magnetic tunnel junctions with a high
tunneling magnetoresistance~\cite{Coey2003,Gross2006}. We have successfully
grown epitaxial thin films of these material
classes~\cite{Reisinger2003a,Philipp2001,Philipp2003a,Majewski2005b,Nielsen2008}
and clarified several physical aspects such as the magnetic
exchange~\cite{Majewski2005a,Majewski2005c,Gepraegs2006,Venkateshvaran2009} or
the influence of strain and steric
effects~\cite{Majewski2005b,Philipp2003b,Philipp2002,Philipp2004}. Here, we
report on the growth and characterization of magnetic oxides and the
integration of such oxides with semiconducting, normal conducting, dielectric
and ferroelectric materials in artificial heterostructures. Special emphasis
will be put on (i) magnetic tunnel junctions consisting of ferromagnetic,
dielectric, and normal conducting layers~\cite{Gross2006}, (ii) transition metal-doped
semiconductors~\cite{Opel2008}, and (iii) multiferroic
materials~\cite{Gepraegs2007}. Along with their integral physical properties
such as electrical conductivity, magnetoresistance, or magnetization we
present data on their microscopic structure with a special emphasis on surface
and interface quality obtained by high-resolution X-ray diffraction techniques
and advanced methods of transmission electron microscopy.

\section{Magnetic Tunnel Junctions Based on Fe$_3$O$_4$.}
 \label{sec:MTJ}

Magnetic tunnel junctions (MTJs) are key elements for the realization of
non-volatile magnetic random access memory (MRAM) devices~\cite{Akerman2005},
magnetic sensors, or programmable logic elements~\cite{Ney2003}. MTJs based on
simple ferromagnetic metals and alloys are well known for many years. The
tunneling magnetoresistive (TMR) effect was first reported in 1975 in
Fe/GeO$_x$/Co MTJs~\cite{Julliere1975}. Here, $\mathrm{TMR} = (R_{\rm
ap}-R_{\rm p})/R_{\rm p}$, with $R_{\rm ap}$ and  $R_{\rm p}$ denoting the
tunnel resistance for the antiparallel and parallel magnetization direction,
respectively. While at room temperature the achieved TMR values could be
increased up to around 10\% in the following years~\cite{Moodera1999}, it was
predicted in 2000 that a further dramatic increase should be possible for
Fe/MgO/Fe MTJs using highly textured MgO(100) tunneling barriers due to the
Bloch state filtering of the electronic wave
functions~\cite{Butler2001,Mathon2001}. Indeed, in 2004 a high TMR effect above
200\% was reported for Fe~\cite{Yuasa2004} or CoFe
electrodes~\cite{Parkin2004}, and today $\mathrm{TMR} > 600$\% is observed in
CoFeB/MgO/CoFeB MTJs at room temperature~\cite{Ikeda2008}.

Very high TMR values are expected for MTJs based on half-metallic ferromagnets
even without wave function filtering, requiring highly textured tunneling
barriers. Along this line, magnetite (Fe$_3$O$_4$) is a promising candidate,
because band structure calculations predict half-metallicity even at room
temperature~\cite{Zhang1991}. Moreover, magnetite has a high Curie temperature
of 860\,K allowing for room temperature applications~\cite{Gorter1955}. Indeed,
spin-resolved photoelectron spectroscopy experiments on magnetite thin films
revealed a spin polarization near the Fermi level of up to 80\% at room
temperature~\cite{Dedkov2002,Fonin2008}. In contrast, the room temperature
TMR effects of magnetite based MTJs are below 20\%. A
summary of recent results is given in Table~\ref{tab:MTJ-TMR_values}
\cite{Aoshima2003,vanderZaag2000,Li1998,Park2003,Matsuda2002,Seneor1999}.
Therefore, it is important to study the behavior of magnetite at the
electrode-barrier interface, the influence of the tunneling barrier itself, and
the magnetic coupling through thin barriers.

\begin{table}
  \centering
  \caption{Room temperature TMR values from various Fe$_3$O$_4$-based MTJs.}
  \begin{tabular}[hb]{ccc|c|c}
    \hline
    Fe$_3$O$_4$ & barrier & counter & TMR & Ref. \\
    orientation & & electrode & (\%) & \\
    \hline
    (001) & MgO & CoO & 0.4 & \cite{vanderZaag2000} \\
    (001) & MgO & CoCr$_2$O$_4$ & 0.5 & \cite{Li1998} \\
    polycrystalline & AlO$_x$ & NiFe & 7 & \cite{Park2003} \\
    (110) & AlO$_x$ & CoFe & 10 & \cite{Matsuda2002} \\
    polycrystalline & AlO$_x$ & Co & 13 & \cite{Seneor1999} \\
    (110) & AlO$_x$ & CoFe/NiFe & 14 & \cite{Aoshima2003} \\
    (001) & AlO$_x$ & Co & 20 & this work\\
    \hline
  \end{tabular}
  \label{tab:MTJ-TMR_values}
\end{table}

We have fabricated MTJs based on epitaxial magnetite thin films.  Five
different materials (MgO, SrTiO$_3$, NdGaO$_3$, SiO$_2$, and AlO$_x$) have been
used for the tunneling barrier. Polycrystalline Ni and Co films deposited by
electron beam evaporation in the same ultra-high vacuum cluster system served
as counter electrodes. According to recent tunneling
experiments~\cite{Moodera1999}, Ni and Co have spin polarizations of 33\% and
42\%, respectively. The magnetic properties of the magnetic heterostructures,
in particular the presence of any magnetic coupling between the ferromagnetic
junction electrodes across the tunneling barrier, was studied by SQUID
magnetometry prior to pattering of the samples. The magnetotransport behavior
of the MTJs was measured as a function of temperature and applied magnetic
field using a standard four probe technique.

\subsection{Fabrication Process.}
 \label{sec:MTJ-PLD}

The thin film heterostructures required for the magnetite based MTJs were
fabricated in an ultra-high vacuum (UHV) cluster system. First, the bottom
electrode consisting of epitaxial TiN and a 20 to 50\,nm thick
Fe$_3$O$_4$ layer were deposited on (001) oriented MgO substrates using
laser-MBE~\cite{Gross2000,Klein1999}. Stoichiometric targets and a KrF excimer
laser (248\,nm) with a fluence of 2\,J/cm$^2$ have been used. The TiN layer was
grown at a substrate temperature of 600$^\circ$C and Fe$_3$O$_4$ was deposited
at 300$^\circ$C, both at an Ar pressure of $3.7 \times 10^{-3}$\,mbar.
The growth was monitored \textit{in-situ} using RHEED. Details
of the deposition process including the infrared laser substrate heating are
given elsewhere~\cite{Gross2000,Klein1999,Reisinger2003a,Reisinger2003b}.

Second, the thin insulating tunneling barrier is
deposited. Five different materials were used: MgO, SrTiO$_3$, and NdGaO$_3$
have been grown by pulsed laser deposition (PLD) from stoichiometric targets,
again in a pure Ar atmosphere of $3.7\times10^{-3}$\,mbar and a substrate
temperature of 330$^\circ$C. SiO$_2$ was deposited by electron beam evaporation
in the same UHV cluster system at room temperature without breaking the vacuum.
The AlO$_x$ tunneling barriers were obtained by evaporating Al also in the same
UHV cluster system and subsequent oxidation in pure O$_2$. The only material
growing epitaxially on Fe$_3$O$_4$ was MgO as confirmed by the corresponding
RHEED pattern.\footnote{Epitaxial growth of MgO by PLD was also achieved at
room temperature in an Ar/O$_2$ mixture (99:1) at $3.4\times10^{-3}$\,mbar.}
The SrTiO$_3$, NdGaO$_3$, SiO$_2$, and AlO$_x$ thin films form polycrystalline
tunneling barriers on Fe$_3$O$_4$.

Third, a 40\,nm thin Ni or Co top electrode serving as the ferromagnetic
counter electrode of the MTJs was deposited \textit{in-situ} by electron beam
evaporation in the UHV cluster system at room temperature. Due to the epitaxial
nature of the MgO tunneling barrier, also epitaxial magnetite counter
electrodes have been used for this barrier type. Finally, for most samples an
\textit{in-situ} Au layer was deposited also by e-beam evaporation to prevent
the oxidation of the ferromagnetic top electrode during the patterning process.
We used an \textit{in-situ} atomic force microscope to probe the surface
quality. The structural quality and layer thicknesses of the
films was determined with a Bruker-AXS high-resolution X-ray diffractometer.

\begin{figure}[tb]
 \centering
    \includegraphics[width=.77\linewidth]{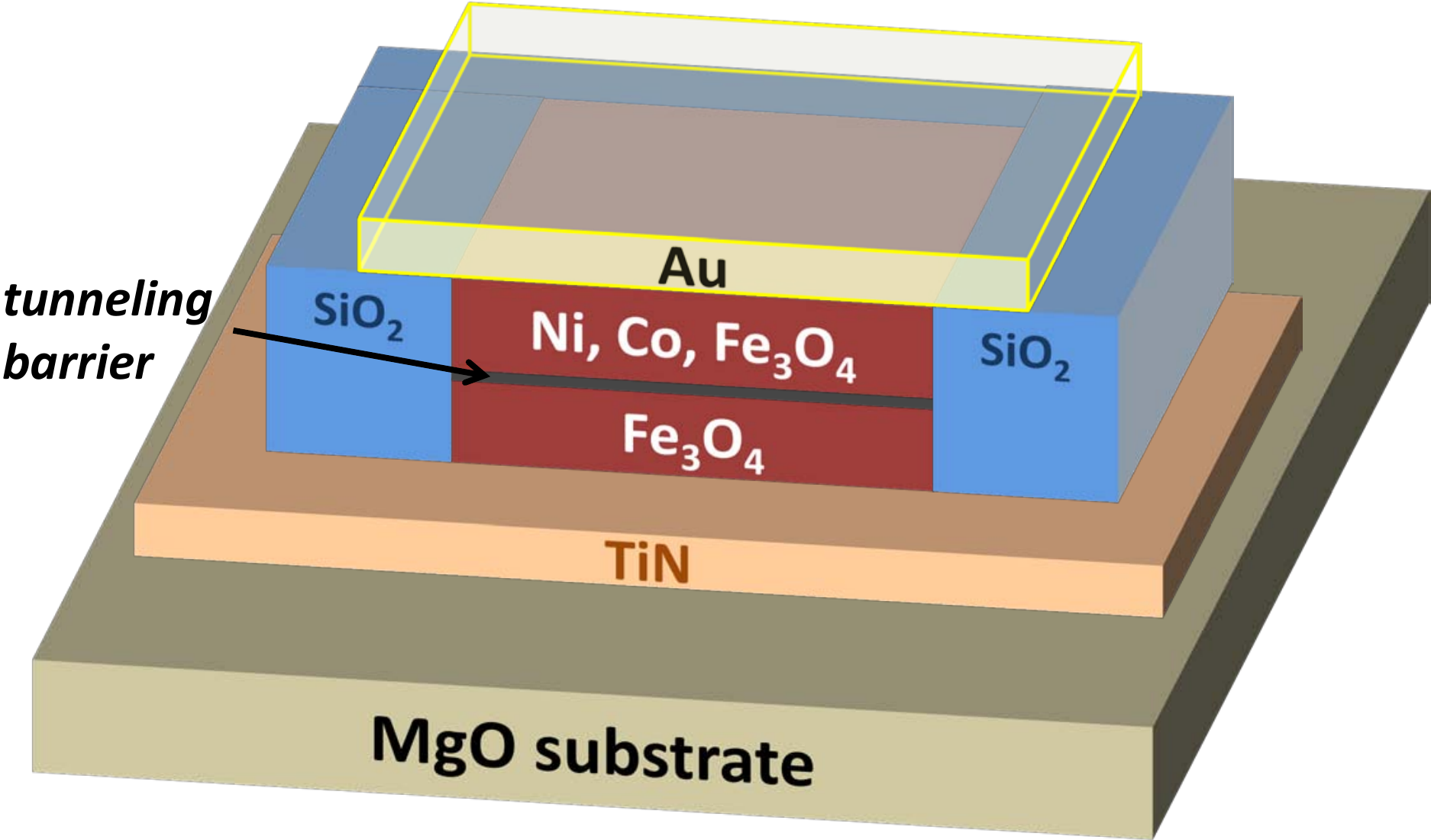}
    \caption{
Sketch of the layer structure of the investigated magnetic tunnel junctions. The
bottom electrode (TiN/Fe$_3$O$_4$) is separated from the top electrode (Ni/Au or Co/Au)
by the thin tunneling barrier (MgO, SrTiO$_3$, NdGaO$_3$, SiO$_2$, or AlO$_x$).
The SiO$_2$ layer provides electrical insulation between the Au wiring of the top electrode
and the bottom electrode (front part of the SiO$_2$ layer removed for clarity).}
    \label{fig:MTJ}
\end{figure}

The multilayer stacks (cf. Fig.~\ref{fig:MTJ}) were used to fabricate MTJs with
different junction areas and shapes by optical lithography and Ar ion beam
milling~\cite{Alff1992}. In a first step, a $100{\rm \mu m}$ wide strip
is patterned into the whole stack. Then, a rectangular or ring-shaped junction
area is defined by etching a mesa structure using an
adequately shaped photoresist stencil as the etching mask. The removed part is
replaced by a polycrystalline SiO$_2$ window insulation, using the same resist
stencil for the lift-off process. Finally, a narrow Au strip is used to provide
the electrical contacts to the top electrode. In this way rectangular MTJs with
areas ranging from $10\times10 {\rm \mu m}^2$ to $5\times100 {\rm \mu m}^2$ and
ring-shaped MTJs with inner (outer) diameters of $6{\rm \mu m}$ ($26{\rm \mu
m}$) and $26.8{\rm \mu m}$ ($36.8{\rm \mu m}$) have been fabricated.

\subsection{Magnetic Characterization of the Junction Electrodes.}
 \label{sec:MTJ-SQUID}

First, a detailed characterization of the magnetite films was performed. The
Fe$_3$O$_4$ thin films were found to show properties which are comparable to
those of single crystal bulk material with respect to crystallographic
structure and magnetization~\cite{Reisinger2004,Venkateshvaran2008}. In
particular, their magnetization versus temperature curves showed a pronounced
\textit{Verwey} transition~\cite{Verwey1939} confirming the high structural
quality and indicating perfect stoichiometry~\cite{Shepherd1985}.

\begin{figure}
 \centering
    \includegraphics[width=\linewidth]{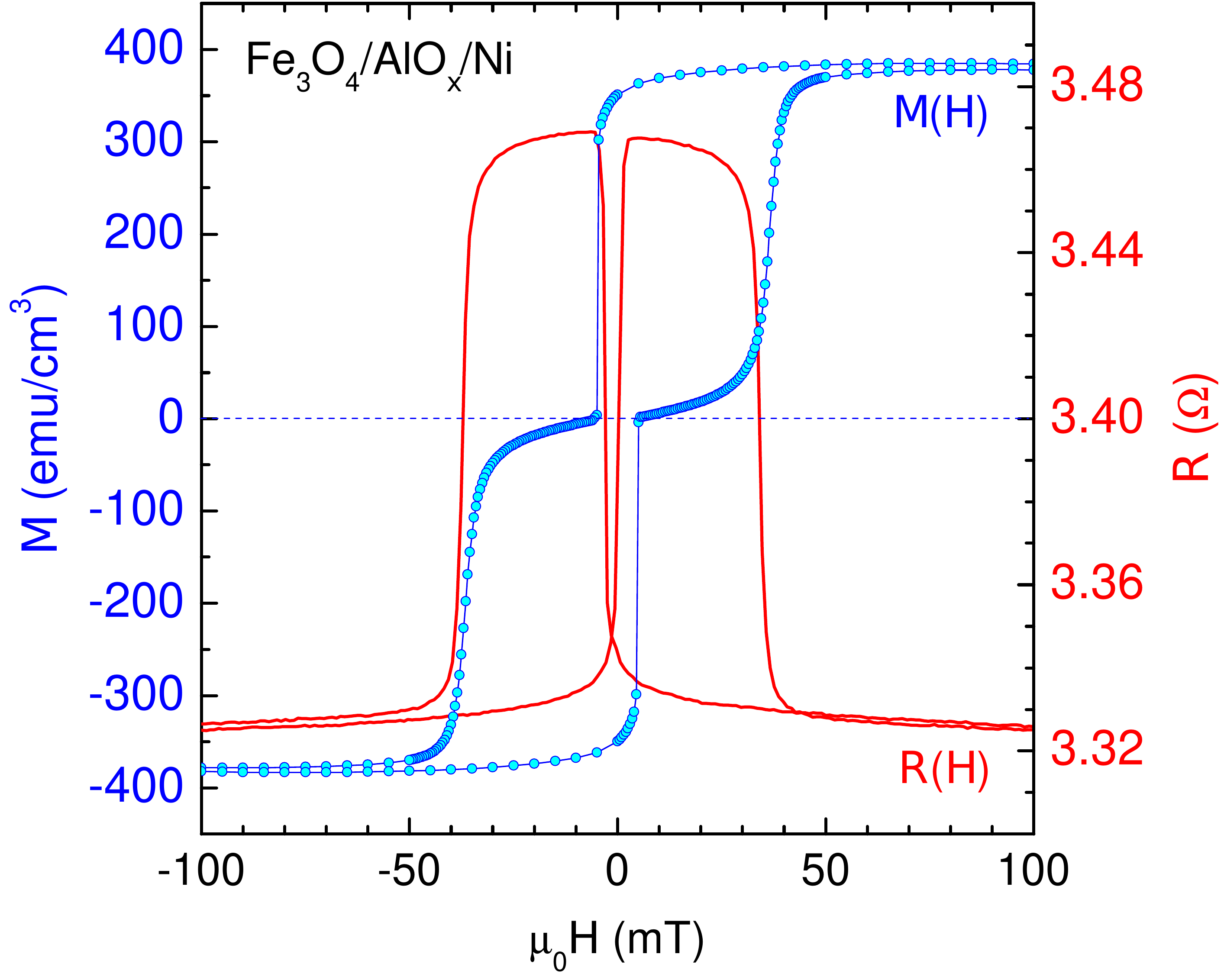}
    \caption{
Magnetization versus applied magnetic field curve of an unpatterned
Fe$_3$O$_4$(52nm)/ AlO$_x$(2.5nm)/ Ni(40nm) stack at 210\,K (blue, left
scale). Also shown is the TMR effect of a $20\times40 {\rm \mu m}^2$ MTJ
patterned into such stack at 320\,K (red, right scale).}
    \label{fig:MTJ-NiMagTMR}
\end{figure}

Next, magnetization versus applied magnetic field curves, $M(H)$, of the
multilayer stacks prior to any photolithographic processing and patterning have
been measured in a SQUID magnetometer at different temperatures. The magnetic
field was applied parallel to the layers. In this way the values of the
coercive field and saturation magnetization in the bottom and counter electrode
can be determined. Furthermore, the presence of a finite magnetic coupling
across the tunneling barrier can be checked. Figure~\ref{fig:MTJ-NiMagTMR}
(blue curve, left scale) shows the $M(H)$ curve of a sample consisting of
14\,nm TiN, 52\,nm Fe$_3$O$_4$, 2.5\,nm AlO$_x$, and 40\,nm Ni at 210\,K.
Obviously, the magnetization of the bottom and counter electrode switches
separately at different fields. From the experimental data the coercivities of
the Ni and the Fe$_3$O$_4$ layer are determined to 5\,mT and 32\,mT at 210\,K,
respectively. These values decrease to 2\,mT and 24\,mT, respectively, at
380\,K (not shown here). Due to the strong difference of the coercive fields an
antiparallel magnetization orientation of the bottom and counter electrode is
achieved over a wide field range. This is a prerequisite for a reliable
determination of the TMR value without using an exchange bias layer.
Furthermore, the clear step-like shape of the $M(H)$ curves (cf.
Fig.~\ref{fig:MTJ-NiMagTMR}, blue) at all temperatures between 210\,K and
380\,K indicates a homogeneous switching of the magnetization direction in the
junction electrodes. This is necessary for achieving an optimum TMR
effect. Both the gradual rotation of magnetic domains and a glass-like magnetic
state would be unfavorable in this respect.

Comparing the coercive field of the Ni counter electrode in
Fig.~\ref{fig:MTJ-NiMagTMR} to that of a 20\,nm thick reference Ni film
(0.25\,mT at 290\,K), we found an increase of about a factor of 10. This effect
is attributed to a residual magnetic coupling through the thin insulating
barrier. Comparing $M(H)$ curves of stacks with different barrier materials, we
find slightly sharper switching steps for SiO$_2$ than for AlO$_x$. This
indicates a more homogeneous switching of the magnetic domains. On the other
hand, the barrier thickness is more difficult to control for SiO$_2$ than for
AlO$_x$. For MgO barriers, the difference between the coercive fields of the
Fe$_3$O$_4$ bottom and the Fe$_3$O$_4$/Ni counter electrode amounts to only about 7\,mT at room
temperature. At lower temperatures (210~K and 150~K) the electrodes even switch
together at approximately 30\,mT. That is, the stack behaves like a single
ferromagnetic layer. This provides strong evidence for a strong coupling of the
ferromagnetic electrodes across the MgO barrier layer. Further details on
TiN/Fe$_3$O$_4$/MgO/Fe$_3$O$_4$/Au MTJs are discussed in
subsection~\ref{sec:MTJ-MgO}. In the following subsection we first focus on
MTJs based on TiN/Fe$_3$O$_4$/AlO$_x$/(Ni;Co)/Au stacks.

\subsection{Magnetotransport Properties.}
 \label{sec:MTJ-MTJ}

The magnetotransport measurements of the MTJs were performed using a standard
four-probe technique in a liquid He cryostat system with variable temperature
insert and a 10\,T superconducting solenoid. For each magnetic field value (or
temperature), the voltage drop across the MTJs was measured with different
polarities of the applied current to eliminate possible thermovoltages. The
magnetic field was always applied in the film plane. We note that the
resistivity of the TiN films is much smaller than that of the magnetite film.
Therefore, this layer serves as a low resistance shunt of the magnetite layer
in the bottom electrode. In the junction area the electrical current is flowing
vertically from the bottom TiN layer through the ferromagnetic bottom electrode
(Fe$_3$O$_4$) across the tunneling barrier to the ferromagnetic top electrode
(Ni or Co). Finally, the current is leaving the junction area via the Au wiring
layer.

\paragraph{Tunneling Resistance.}

Information on the on-chip homogeneity of the tunneling barrier can be obtained
by comparing the resistance times area product, $R \cdot A$, of junctions
patterned on the same chip. In Table~\ref{tab:MTJ-RA} we have listed the room
temperature $R \cdot A$ values of junctions with different area $A$ but the
same thickness ($\sim 2.5$\,nm) of the AlO$_x$ tunneling barrier. As expected for a
spatially homogeneous barrier, we find $R \propto 1/A$ or, equivalently, $R
\cdot A \sim const.$ We found that the junctions with higher
$R \cdot A$ values show a significantly larger TMR \cite{Chen2000}.
We note that the $R \cdot A$ values found for the AlO$_x$ barrier in our
experiment are lower than those observed by other groups \cite{Gao2009}, but
comparable to those reported for a lower barrier thickness \cite{Chen2000}.
This can be explained by assuming a smaller oxide thickness in our AlO$_x$ or
may point to the presence of defect states in our barrier, as suggested by the
$R(T)$ behavior which will be discussed in the following.

\begin{table}
  \centering
  \caption{Resistance$\times$area product, $R \cdot A$, of rectangular-shaped MTJs
  with AlO$_x$ barrier (thickness $\sim 2.5$\,nm) at room temperature.}
  \begin{tabular}{cc|c}
    \hline
    area $A$ (${\rm \mu m}^2$) & $R$ ($\rm \Omega$) & $R{\cdot}A$ (${\rm \Omega m}^2$)\\
    \hline
    $10\times10$ & 66 & $6.6\cdot10^{-9}$ \\
    $10\times20$ & 36 & $7.2\cdot10^{-9}$ \\
    $20\times20$ & 23 & $9.2\cdot10^{-9}$ \\
    \hline
  \end{tabular}
  \label{tab:MTJ-RA}
\end{table}

\begin{figure}
 \centering
    \includegraphics[width=\linewidth]{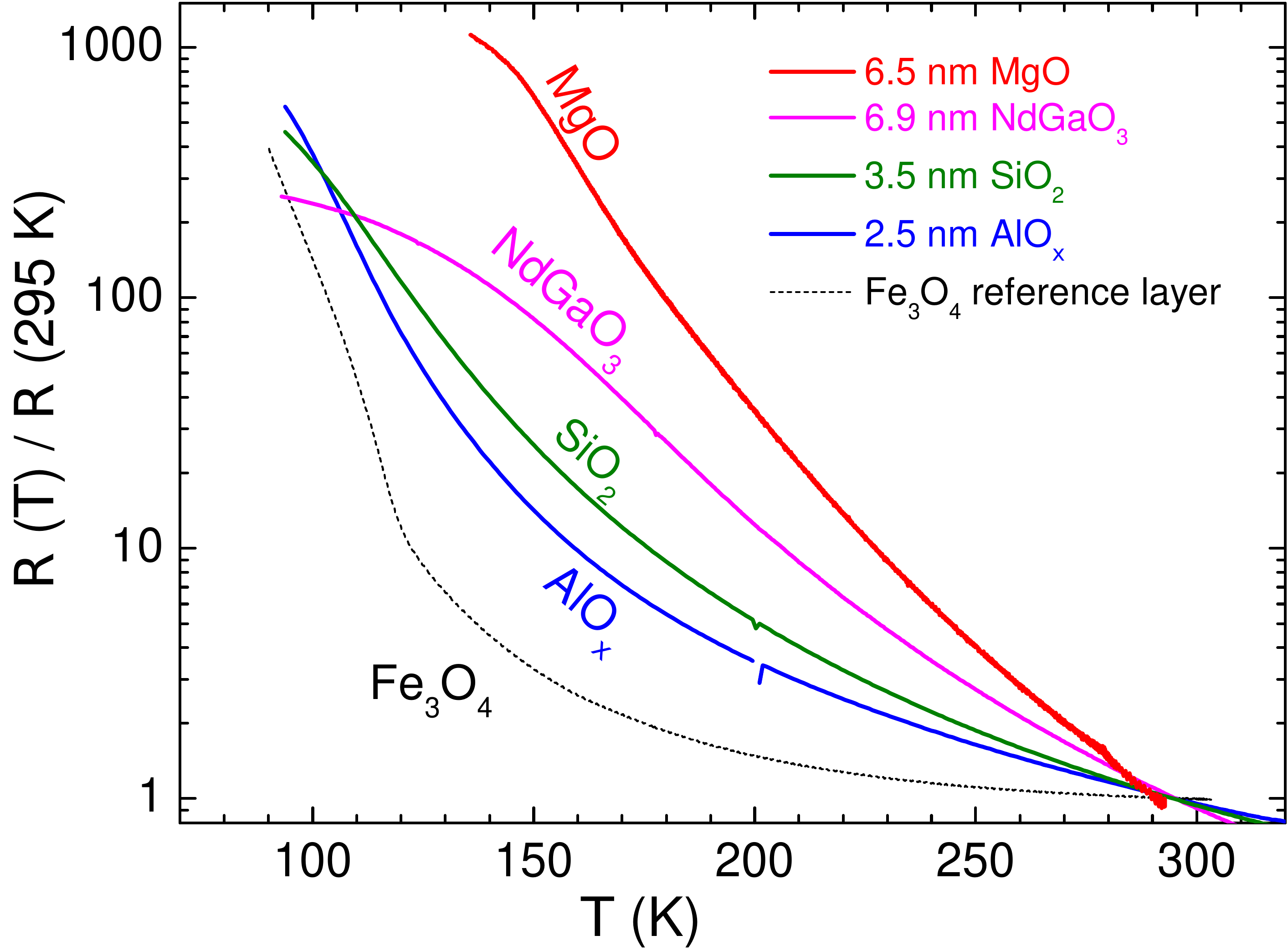}
    \caption{Comparison of the temperature dependence of the junction resistance, $R(T)$,
    of MTJs with different tunneling barriers. The resistance values are normalized
    to the room temperature value at 295\,K. }
    \label{fig:MTJ-rho}
\end{figure}

Figure~\ref{fig:MTJ-rho} shows normalized resistance versus temperature curves,
$R(T)/R(295\,\textrm{K})$, obtained for MTJs with five different barrier
materials. For comparison we also show the normalized $R(T)$ curve of a
magnetite thin film. The absolute resistance values vary between about 5\,$\rm
\Omega$ to $10\,{\rm k\Omega}$ at room temperature, depending on the barrier
type and junction area. Due to the Verwey transition at $T_{\rm V} \simeq
120$\,K \cite{Verwey1939}, the resistance of the magnetite layer increases
strongly below this temperature.

Figure~\ref{fig:MTJ-rho} clearly shows a strong increase of the junction
resistance with decreasing temperature. In contrast, for an ideal tunneling
barrier with barrier higher $eV_B$ much larger than $k_BT$ we expect only a
slight increase of the tunneling resistance $R$ due to the reduction of the
smearing of the Fermi distribution with decreasing temperature. This indicates
that all tunneling barriers are far from being ideal and most likely have a
considerable density of defect states mediating thermally activated inelastic
tunneling. According to the data of Fig.~\ref{fig:MTJ-rho}, this effect is
strongest for the MgO and weakest for the AlO$_x$ barrier, suggesting that the
MgO barrier has the highest and the AlO$_x$ the lowest density of defect
states. We note that the MgO, SrTiO$_3$, and NdGaO$_3$ barriers are
deposited by PLD from stoichiometric targets in pure Ar atmosphere to avoid
oxidation of the Fe$_3$O$_4$ base electrode. This is known to result in oxygen
deficiency and, in turn, defect states in the barrier layer. The AlO$_x$
barrier, however, is obtained by evaporation of Al and subsequent oxidation in pure
oxygen atmosphere. As pointed out above, for the MTJs with the MgO barrier
magnetic coupling of the junction electrodes is observed despite the large
barrier thickness of 6.5\,nm. This coupling may be mediated by the high density
of defect states in the barrier together with an Fe doping of the barrier
layer by interdiffusion. Although this scenario is quite likely, a more
detailed study is required to clarify this issue.

We also have to address the contribution of the magnetite layer to the measured
junction resistance. Assuming a homogeneous current feeding into the junction
through the well conducting TiN bottom layer, the resistance of the
magnetite layer for the current perpendicular to plane configuration is
estimated to  $5.6\,{\rm m\Omega}$ at room temperature for a junction area of
$20\times20 {\rm \mu m}^2$. Here, we use the measured resistivity and thickness
of the magnetite layer. Comparing this value with the total junction resistance
we see that the additional resistance of the magnetite layer is negligible in
most cases, at least at temperatures well above the Verwey transition
temperature. As will be pointed out below, there are situations where the influence of the
magnetite electrode has to be taken into account. In this case, unusually high
effective TMR values can be obtained due to current redistribution.

\paragraph{Tunneling Magnetoresistance.}

From the measured $R(H)$ curves we have determined the TMR effect
\begin{equation}
  \textrm{TMR} (H) = \frac{R_{\rm ap} - R(H)}{R_{\rm p}}
  \label{eq:TMR}
\end{equation}
for MTJs with different barrier materials in the temperature regime between
150\,K and 350\,K. Here, $R_{\rm ap}$ is the resistance corresponding to the maximum resistance of the
respective MTJ. $R_{\rm p}$ is the resistance for the parallel magnetization
configuration. For the evaluation of $\textrm{TMR}(H)$ we always used the
junction resistance measured at $+300$\,mT for $R_{\rm p}$. All investigated
samples reproducibly show a positive TMR effect. Furthermore, the
$\textrm{TMR}(H)$ curves are symmetric with respect to $H=0$ and show a sharp
switching behavior at the coercive fields. A typical $R(H)$ curve of a
Fe$_3$O$_4$(52\,nm)/AlO$_x$(2.5\,nm)/Ni(40\,nm) MTJ is shown in
Fig.~\ref{fig:MTJ-NiMagTMR} (red, right scale). A pronounced TMR effect is
observable in the whole investigated temperature range from 150\,K to 350\,K.
The switching fields correspond to the coercive fields of the ferromagnetic
junction electrodes (cf. Fig.~\ref{fig:MTJ-NiMagTMR}, blue). We note that the
increase of resistance for antiparallel electrode magnetization (positive TMR
effect) provides clear evidence for a negative spin polarization of Fe$_3$O$_4$
as the ferromagnetic Ni counter electrode has a negative spin polarization. For
an electrode with positive spin polarization such as the colossal
magnetoresistive manganite La$_{0.7}$Sr$_{0.3}$MnO$_3$, a negative TMR effect
is expected and was indeed reported in literature~\cite{deTeresa1999,Hu2002}.

\begin{figure}
 \centering
    \includegraphics[width=\linewidth]{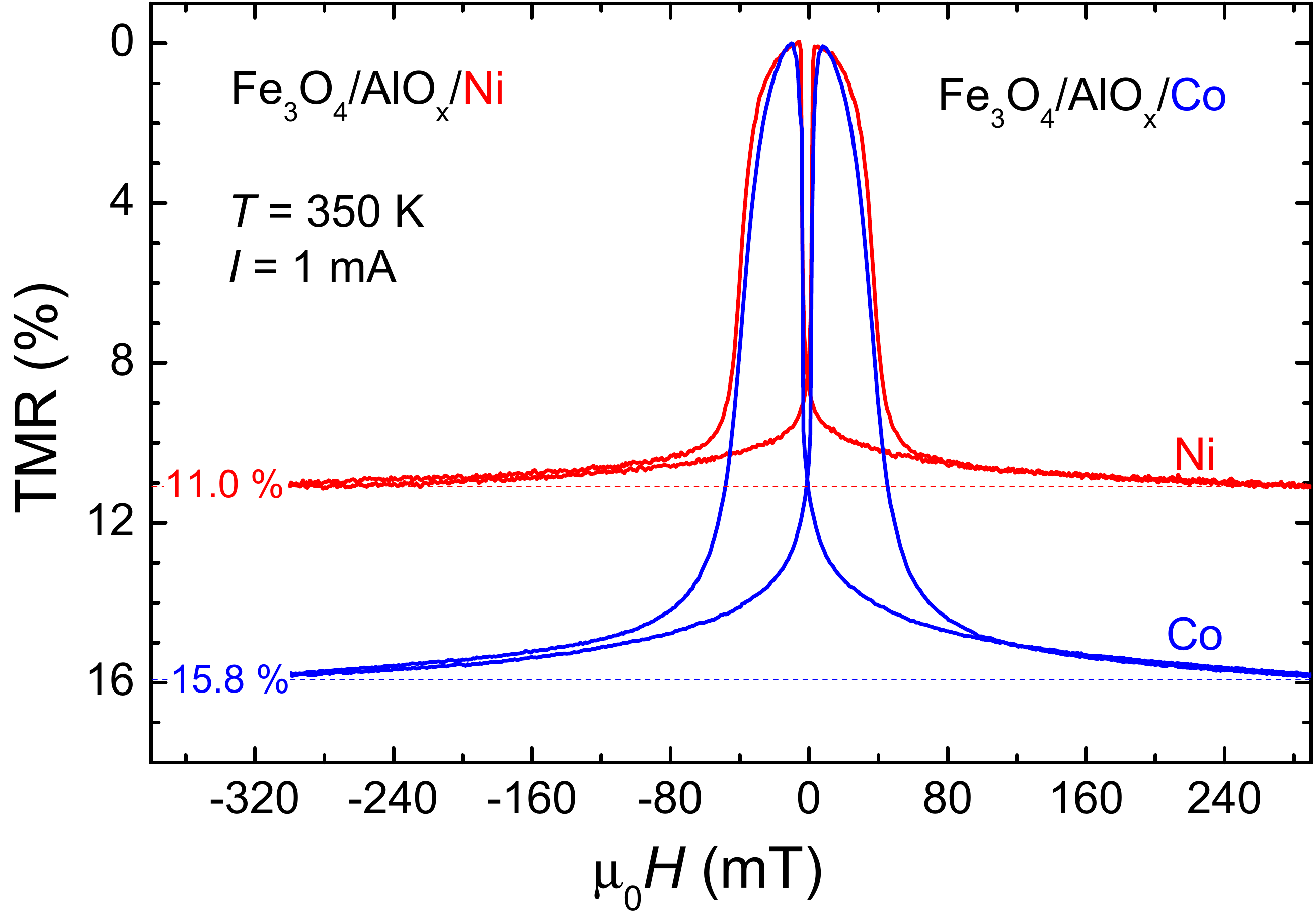}
    \caption{
$\textrm{TMR}(H)$ curves obtained at 350\,K for two MTJs with a Fe$_3$O$_4$
base electrode, an AlO$_x$ tunneling barrier, and a Ni (red) or Co (blue)
counter electrode. }
    \label{fig:MTJ-NiCoTMR}
\end{figure}

In the following, we obtain TMR values of magnetite-based MTJs with different counter
electrodes (Ni and Co) and we estimate the spin polarization of the
Fe$_3$O$_4$ electrode. As shown in Fig.~\ref{fig:MTJ-NiCoTMR}, MTJs with rectangular shape having the
same Fe$_3$O$_4$ base electrode and same AlO$_x$ tunneling barrier, but a Ni or
Co counter electrode, show a maximum TMR effect of 11\% for Ni and 15.8\% for Co
at 350\,K. A different sample with an improved ring-shaped geometry revealed
a maximum TMR effect of 20\% at room temperature and 27\% at 350\,K for Co (not shown here).
To our knowledge, the latter value is larger than the highest TMR
effects reported so far for Fe$_3$O$_4$ based MTJs at or above room temperature
(cf.~Table~\ref{tab:MTJ-TMR_values}). Using the simple Julli\'{e}re
model~\cite{Julliere1975}, the measured TMR values can be related to the spin
polarizations $P_1$ and $P_2$ of the ferromagnetic electrodes according to
\begin{equation}
    \textrm{TMR} = \frac{2P_1P_2}{1-P_1P_2} \;\; .
    \label{eq:Julliere}
\end{equation}
With the literature values $P_{\rm Co} = -42\%$ and $P_{\rm Ni} = -33\%$ \cite{Moodera1999},
we derive a spin polarization for magnetite at room temperature between $P_{\rm Fe_3O_4} \simeq -16\%$ (Ni)
and $P_{\rm Fe_3O_4} \simeq -28\%$ (Co). However, this value is by far too
low as compared to recent data ($-80\%$) obtained by
spin-resolved photoelectron spectroscopy~\cite{Dedkov2002,Fonin2008}.
We attribute this to spin-flip scattering at
defect states in the tunneling barrier or inelastic tunneling processes
involving magnon scattering. For example, if a finite amount of the tunneling
current is involving spin-flip processes, it has been shown that the ideal TMR
value of eq.(\ref{eq:Julliere}) is reduced to~\cite{Hoefener2000,Klein1999a}
\begin{equation}
    \mathrm{TMR}^\star = \mathrm{TMR} \; \left( 1- \frac{I_{\rm ap}^{\rm sf}}{I_{\rm ap}^{\rm tot}} \right) \;\; .
    \label{eq:spinpol_b}
\end{equation}
Here, $I_{\rm ap}^{\rm sf}$ is the part of the tunneling current involving
spin-flip processes and $I_{\rm ap}^{\rm tot}$ is the total tunneling current
in the antiparallel magnetization configuration. There may be various sources
of $I_{\rm ap}^{\rm sf}$ such as inelastic tunneling via defect states in the
barrier or tunneling processes involving magnon scattering in the junction
electrodes. It is immediately evident that a finite value of $I_{\rm ap}^{\rm
sf}$ is causing a reduction of the measured TMR effect. The strong temperature
dependence of the tunneling resistance of our junctions (cf.
Fig.~\ref{fig:MTJ-rho}) shows that there is a large amount of inelastic
tunneling current. This is consistent with the small absolute TMR$^\star$
values of up to 27\% measured for our investigated MTJs.

The determination of the sign of $P_{\rm Fe_3O_4}$ cannot be done unambiguously
by evaluating the TMR effect in magnetic tunnel junctions
as it depends on the sign of $P$ of the counter electrode.
From theory, $P_{\rm Ni,Co} < 0$ is predicted and observed for Ni in spin-polarized scanning
tunneling microscopy experiments \cite{Alvarado1992}, but not in electrical
transport at AlO$_x$ interfaces \cite{Moodera1999}. From eq.~(\ref{eq:Julliere}),
$P_{\rm Ni,Co} < 0$ leads to a negative spin polarization for Fe$_3$O$_4$
as discussed above and reported by spin-resolved photoelectron spectroscopy
\cite{Dedkov2002,Fonin2008}, even through an AlO$_x$ barrier \cite{Bataille2007}.
However, assuming $P_{\rm Ni,Co} > 0$ results in a positive value for $P_{\rm Fe_3O_4}$
and is reported for transport across the AlO$_x$ interface by the same authors \cite{Bataille2007}.

\subsection{Giant TMR Effect by Current Redistribution.}

\begin{figure}
 \centering
    \includegraphics[width=\linewidth]{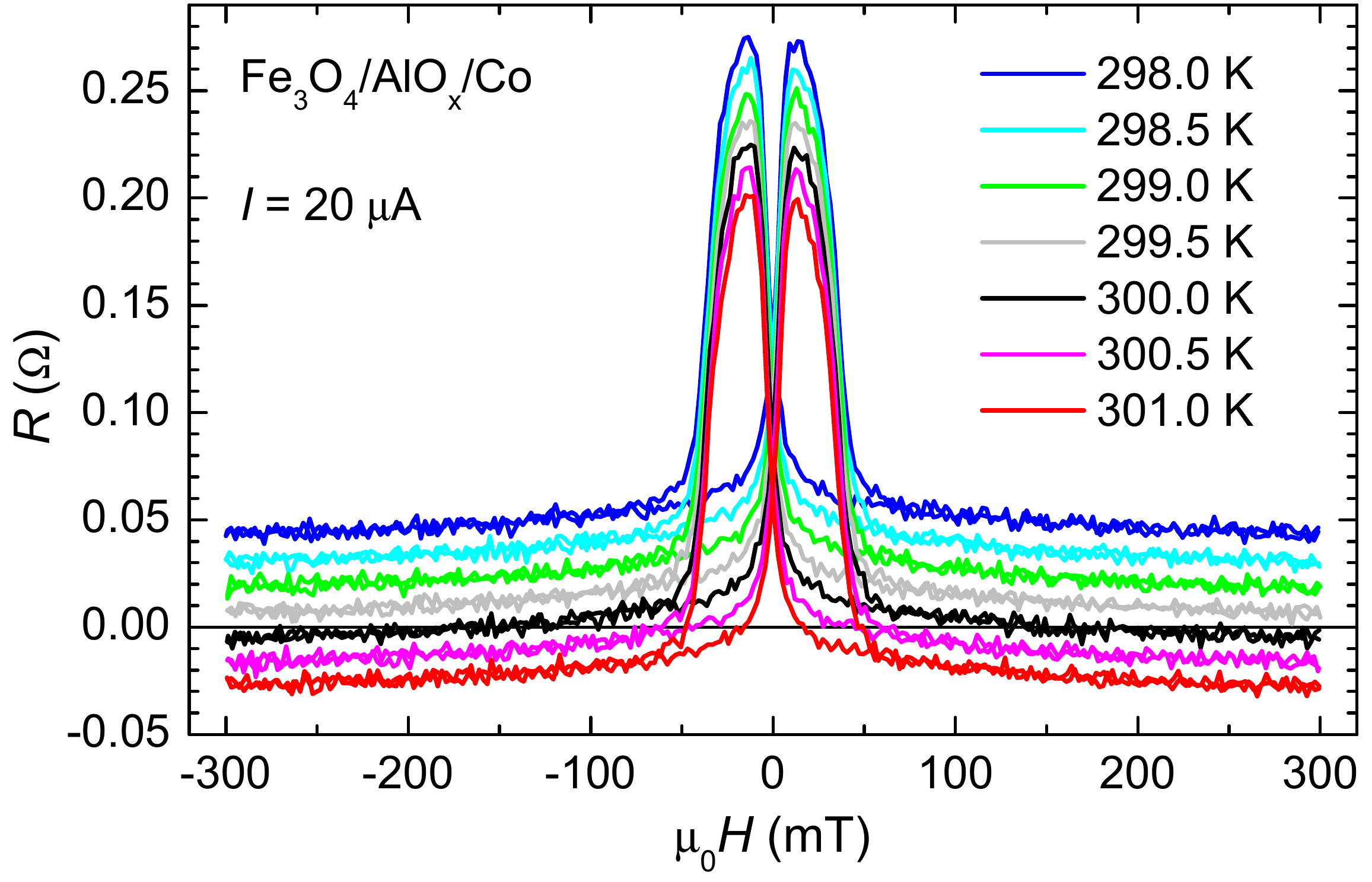}
    \caption{
Resistance versus applied magnetic field curves, $R(H)$, of a ring-shaped MTJ
(inner diameter: $26.8\,{\rm \mu m}$, outer diameter: $36.8\,{\rm \mu m}$) at
different temperatures. At 299.5\,K, the $R(H)$ curve approaches zero for
parallel magnetization, resulting in an extremely high TMR value due to $R_{\rm
p} \rightarrow 0$.}
    \label{fig:MTJ-Edwin}
\end{figure}

A detailed study of MTJs with different shape of the junction area showed that
for specific samples the TMR values derived according to eq.~(\ref{eq:TMR}) can
become huge, change sign, and depend sensitively on temperature. This effect
could be reproduced in several samples. However, the temperature regime in
which this effect occurs strongly depends on the geometry and the layer
structure of the MTJs. To illustrate this effect, in Fig.~\ref{fig:MTJ-Edwin}
we show $R(H)$ curves of a ring-shaped MTJ based on a
TiN(35\,nm)/Fe$_3$O$_4$(20\,nm)/AlO$_x$ (3\,nm)/Co(15\,nm) stack at different
temperatures around $T = 300$\,K. The resistance $R(H)$ is obtained by simply
dividing the measured voltage drop $V$ across the MTJ by the applied current of
$I=20\,{\rm \mu A}$. Obviously, the $R(H)$ curves are shifted to smaller values
with increasing temperature. Most interestingly, for parallel magnetization of
the ferromagnetic layers, i.e. for applied fields well above the coercive
fields of the electrode materials, the $R(H)$ curve is close to zero and even
becomes negative with increasing temperature. Here, negative $R$ means that the
measured voltage drop across the junction has become negative. If we use the
$R(H)$ curves of Fig.~\ref{fig:MTJ-Edwin} to formally derive a TMR value
according to eq.~(\ref{eq:TMR}), we obtain giant values of several 1000\%,
since we divide by a very small $R_{\rm p}$ value. Furthermore, the derived TMR
value can change sign due to the sign change of $R_{\rm ap}$. We note, however,
that the overall difference $R_{\rm ap}-R_{\rm p}$ between the anti-parallel
and the parallel magnetization configuration of the MTJ does not change
significantly.

We attribute the effect described above to the specific choice of the junction
geometry and layer structure. By a proper choice of parameters a significant
redistribution of the current in the junction electrodes is achieved. A similar
effect has been reported earlier~\cite{Moodera1996} for junctions with very
large area and small thickness of the junction electrodes. When using very thin
or highly resistive junction electrodes, the voltage drop along the junction
electrodes can become comparable to the voltage drop across the tunneling
barrier. To derive the TMR value, only the voltage drop across the tunneling
barrier is required. However, in the real experimental situation both this
voltage drop and part of the voltage drop along the junction electrodes is
measured. In this case, the determination of the TMR value is difficult and a
naive analysis can lead to giant, geometry-induced TMR effects as discussed
above. Whereas for junctions with metallic electrodes this effect only appears
for MTJs with very large junction area and thin electrodes, in our MTJs this
effect can be reproducibly obtained also for small junction dimensions in the
micrometer range and therefore may be interesting for applications. This is
caused by the much higher resistivity of the base electrode of our MTJs.

\begin{figure}
 \centering
    \includegraphics[width=\linewidth]{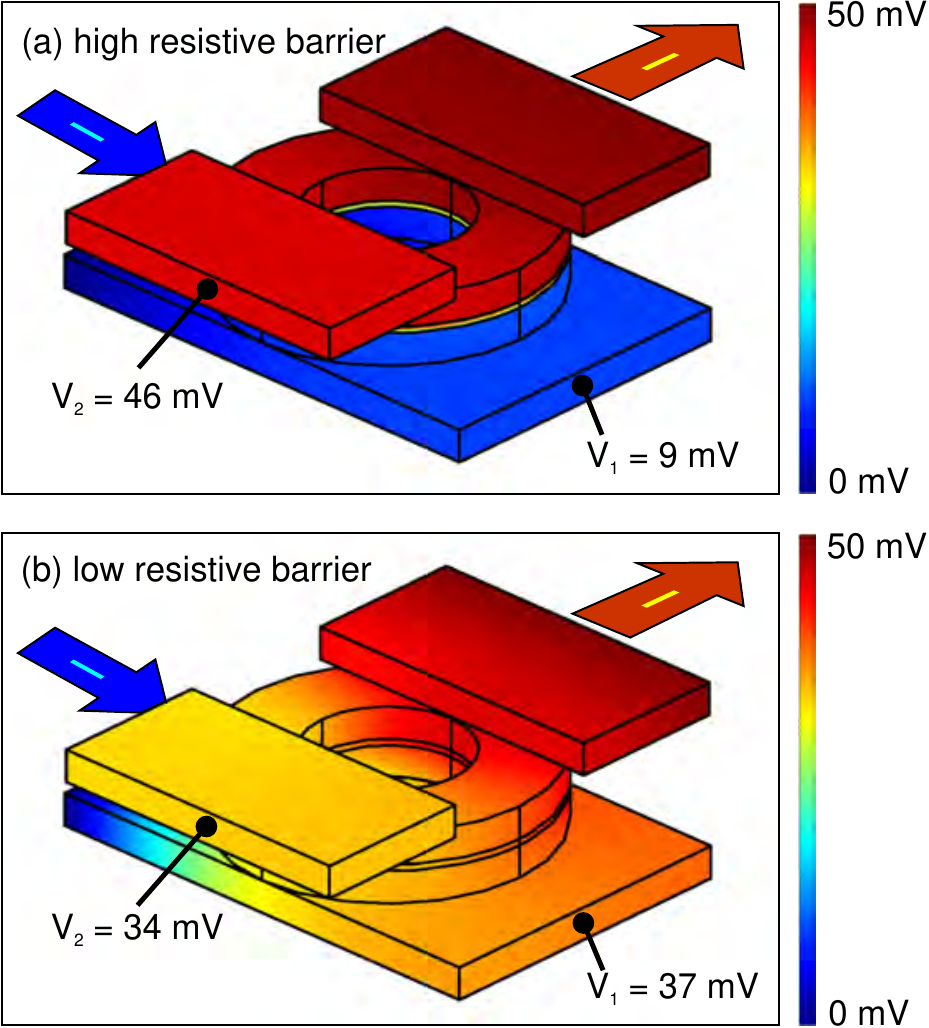}
    \caption{
FEM simulations of the electrical potential distribution in ring-shaped MTJs.
In (a) the situation for a high, in (b) the situation for a low value of the
tunneling resistance is shown. The color code shows the electrical potential
distribution.}
    \label{fig:MTJ-simu}
\end{figure}

To confirm the scenario discussed above, we calculated the spatial distribution
of the electrical potential and the current in ring-shaped MTJs using finite
element method (FEM) simulations. The geometry of the ring-shaped junctions is
shown in Fig.~\ref{fig:MTJ-simu}. The in-plane conductivities used in the simulations
for the various layers of the junction stack are summarized in
Table~\ref{tab:MTJ-simu}.
Since the simulation tool did not allow to represent the real dimensions of the sample because of its large aspect ratio
we scaled the geometry and the conductivity in the out-of-plane direction by a factor of 100.
Fig.~\ref{fig:MTJ-simu} shows the calculated
electrical potential distribution in the MTJ. The current $I$ is driven from the
lower left to the upper right by a fixed source voltage of
50\,mV. The voltage drop across the MTJ is measured between the lower right
($V_1$) and the upper left ($V_2$).

\begin{table}
  \centering
  \caption{Electrical conductivities (in-plane values) of the various layers of the junction stack
           used for the FEM simulations.}
  \begin{tabular}{c|c}
    \hline
    material & conductivity $\sigma$ $({\rm \Omega m})^{-1}$\\
    \hline
    bottom electrode, rectangular (TiN)    & $9.1\times10^6$\\
    bottom ferromagnet, ring (Fe$_3$O$_4$) & $2.25\times10^4$\\
    tunneling barrier, ring                & $1.0\times10^{2\ldots4}$\\
    top ferromagnet, ring (Co)             & $1.79\times10^7$\\
    wiring layer, rectangular (Au)         & $4.85\times10^7$\\
    \hline
  \end{tabular}
  \label{tab:MTJ-simu}
\end{table}

Figure~\ref{fig:MTJ-simu}(a) shows the simulation result for a highly resistive
barrier ($\sigma = 10^2\,{\rm \Omega^{-1}m^{-1}}$). The conductivity of the
tunneling barrier is lower by more than two orders of magnitude as compared to the conductivities
of all other layers. This is the usual situation present in most MTJs with high
conductivity metal electrodes. The color-coded representation of the potential
distribution immediately shows that in this case almost the whole voltage drop
occurs across the barrier, whereas the potential distribution in the bottom
electrode (blue) and the top electrode (red) is homogeneous. From the
simulation we obtain $V_1 = 9$\,mV and $V_2 = 46$\,mV, resulting in a voltage
drop $V_2 - V_1 = 37$\,mV between the voltage probes. Fig.\ref{fig:MTJ-simu}(b)
shows the simulation result for a tunneling barrier with higher
conductivity ($\sigma = 10^4\,{\rm \Omega^{-1}m^{-1}}$). Now the conductivity
of the tunneling barrier is of the same order as the conductivity of the bottom
electrode. In this case a completely different potential distribution is
obtained. Only a minor part of the voltage drop occurs across the barrier and
an inhomogeneous potential distribution is obtained within the junction area.
The calculated potentials at the position of the voltage probes are $V_1 =
37$\,mV and $V_2 = 34$\,mV, resulting in a \emph{negative} voltage drop of $V_2
- V_1 = -3$\,mV between the voltage probes. This result even holds when
increasing the width and length of the bottom electrode. The FEM simulations
clearly show that the measured voltage drop $V_2 - V_1$ can become very small
and even negative if the total resistance of the tunneling barrier becomes
smaller than the total resistance of the electrode material in the junction
area. For our MTJs this is the case due to the low conductivity of the bottom
electrode. For MTJs with high conductivity metal electrode, this is the case
only for very large junction area and/or very thin junction electrodes.

We note that an equivalent effect is known for Josephson junctions in the
normal state, if the current leads have resistivities of the same order as the
insulating barrier. For a structure where the top and bottom electrodes are
parallel and separated by a rectangular insulating barrier, Pedersen \textit{et
al.} developed a one-dimensional model~\cite{Pedersen1967}. For MTJs with a so-called
cross-strip geometry where the top and bottom electrodes are aligned
perpendicular to each other, the current distribution was investigated using
two-dimensional FEM simulations by van de Veerdonk \textit{et
al.}~\cite{Veerdonk1997}. They showed that an inhomogeneous current
distribution across tunnel junctions can suppress the measured voltage drop
which may result in a divergence of the determined TMR value. This is in good
agreement with the results of our three-dimensional FEM simulations describing
the more complex geometry of our MTJs. We note that our simulations have been
performed as proof of principle and are not intended to reproduce all details
of the real junctions.

\subsection{Fully Epitaxial MTJs: The Case of the MgO Barrier.}
 \label{sec:MTJ-MgO}

According to the high spin polarization of magnetite, very high TMR values are
expected for MTJs consisting of two Fe$_3$O$_4$ electrodes. Therefore, we have
grown fully epitaxial TiN/Fe$_3$O$_4$/MgO/Fe$_3$O$_4$/Co/Au stacks on (001)
oriented MgO substrates as described in~\ref{sec:MTJ-PLD}. A thin Co layer on
top of the ferromagnetic Fe$_3$O$_4$ counter electrode was introduced to modify
the coercive field of the counter electrode. In this way different coercive
fields of the bottom and top electrode could be achieved, allowing for the
realization of the antiparallel magnetization configuration. However, the
magnetic characterization of the stacks showed a magnetic coupling of the two
ferromagnetic electrodes at temperatures below 250\,K. As already discussed
in~\ref{sec:MTJ-SQUID}, this coupling results in a synchronous switching of the
magnetization in the two electrodes despite the different coercivities.
Therefore, a stable antiparallel magnetization configuration of the two
electrodes cannot be achieved. There are different scenarios for the origin of
the coupling. A simple origin of ferromagnetic coupling are pinholes in the
barrier layer. However, the TEM analysis of the MTJs showed that the MgO
barriers are homogeneous and pinholes in the typically 5\,nm thick barrier
layer are unlikely. Another possibility are Fe impurities and a large density
of defect states within the MgO barrier which mediate a ferromagnetic coupling
at low temperatures. The impurities may stem from the high energy Fe ions in
the plasma plume, impinging on the barrier layer during the deposition of the
Fe$_3$O$_4$ counter electrode thereby causing intermixing. To clarify this
issue, a combined transmission electron microscopy (TEM) and electron energy-loss
spectroscopy (EELS) investigation was performed.

\begin{figure}
    \centering
    \includegraphics[width=\linewidth]{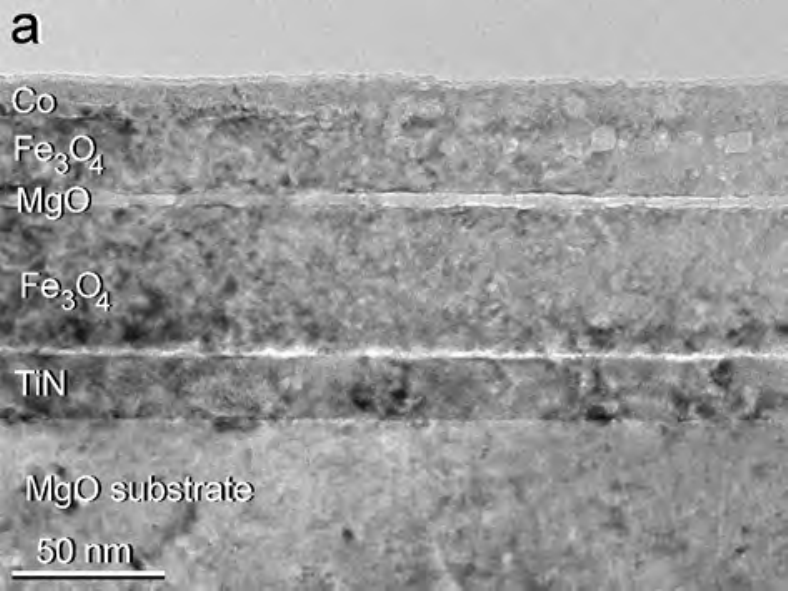}
    \includegraphics[width=\linewidth]{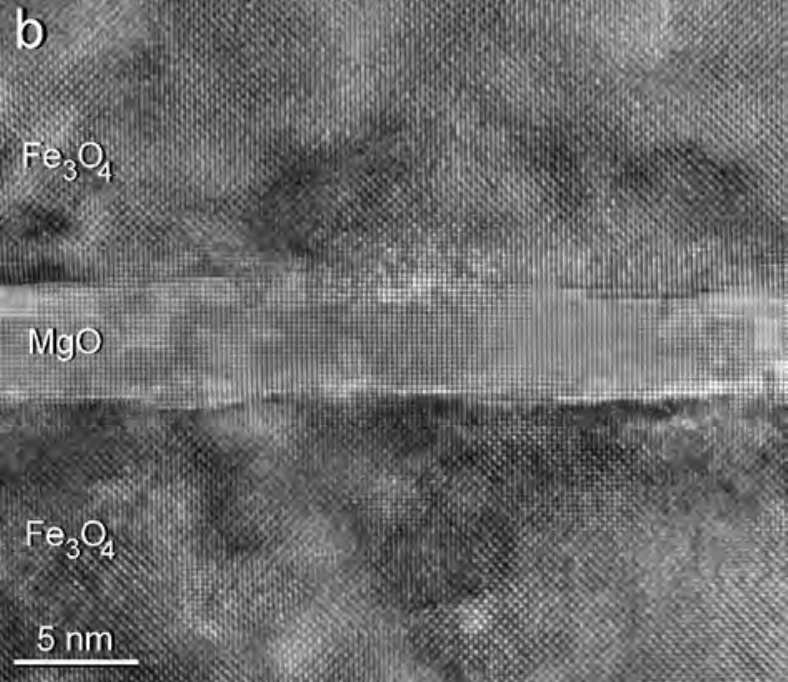}
    \caption{
Transmission electron micrographs of a MTJ stack consisting of TiN (21\,nm),
Fe$_3$O$_4$ (46\,nm), MgO (5\,nm), Fe$_3$O$_4$ (22\,nm), and Co (8\,nm) grown
on a MgO(001) substrate. (a) Bright field image of the complete MTJ stack. (b)
HR-TEM image of the MgO tunnel barrier between the Fe$_3$O$_4$ layers showing
epitaxial growth of all the layers, i.e. the oxygen sub-lattice continues
across the interfaces into the layers. Image in [100] direction of Fe$_3$O$_4$
and of MgO.}
    \label{fig:MTJ-TEM}
\end{figure}

A TEM micrograph of the MTJ stack in cross-section is shown in
Fig.~\ref{fig:MTJ-TEM}(a). The layers of the stack were measured to 21\,nm
(TiN), 55\,nm (Fe$_3$O$_4$), 5\,nm (MgO), 22\,nm (Fe$_3$O$_4$), and 8\,nm (Co).
The thin isolating MgO barrier layer as well as the Fe$_3$O$_4$
electrodes can be seen to be grown homogeneously with constant thickness. The
HR-TEM image in Fig.~\ref{fig:MTJ-TEM}(b) shows that the MgO barrier exhibits
some thickness variations which are the result of steps and ledges at the two
interfaces between MgO and the adjacent Fe$_3$O$_4$ layers. Nevertheless, at
some positions without steps a clear interface can be observed at the two
oxides proving an atomically sharp transition from Fe$_3$O$_4$ spinel to MgO.
The continuation of the (200) lattice planes of MgO and the (400) planes of
Fe$_3$O$_4$, running vertically, prove perfect epitaxy between the lower
magnetite layer across the MgO barrier to the upper magnetite layer. This
observation definitely proves the continuation of the oxygen sub-lattices in
the three oxide layers.

\begin{figure}
    \centering
    \includegraphics[width=\linewidth]{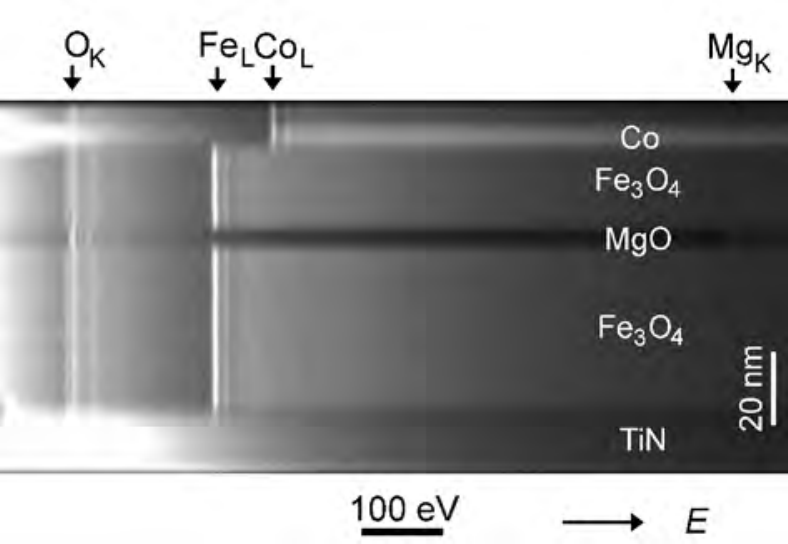}
    \caption{
Spectroscopic image of the MTJ stack in Fig.~\ref{fig:MTJ-TEM} consisting of
EEL spectra (horizontal coordinate: energy loss $E$) as function of position
(vertical coordinate) across the MTJ stack from TiN (bottom) to the Co layer
(top).}
    \label{fig:MTJ-Linescan}
\end{figure}

To investigate the chemical composition of the MgO tunneling barrier, EELS
was applied with an electron probe of less than
1\,nm in diameter. The beam was scanned with 1\,nm increment across the layer
stack, and EEL spectra were acquired simultaneously at any beam position. This
technique, named energy-loss spectroscopic profiling (ELSP)
\cite{Kimoto1997,Walther2003}, results in spectroscopic images with EELS
intensity as a function of energy loss $E$ and the spatial co-ordinate of the
electron beam line profile as shown in Fig.~\ref{fig:MTJ-Linescan}. In the
energy loss range shown the spectrum image contains the $L$ edges of Co and Fe
where the MgO barrier causes the gap in the Fe-$K$ EELS signal. The Mg-$K$ edge
is at $E=1305$\,eV and is thus of low intensity. The oxygen-$K$ edge reveals
the difference in near-edge structure of the MgO and magnetite layers,
respectively, which is the result of the different chemical neighbors of oxygen
and of the differing co-ordination.

\begin{figure}[b]
    \centering
    \includegraphics[width=\linewidth]{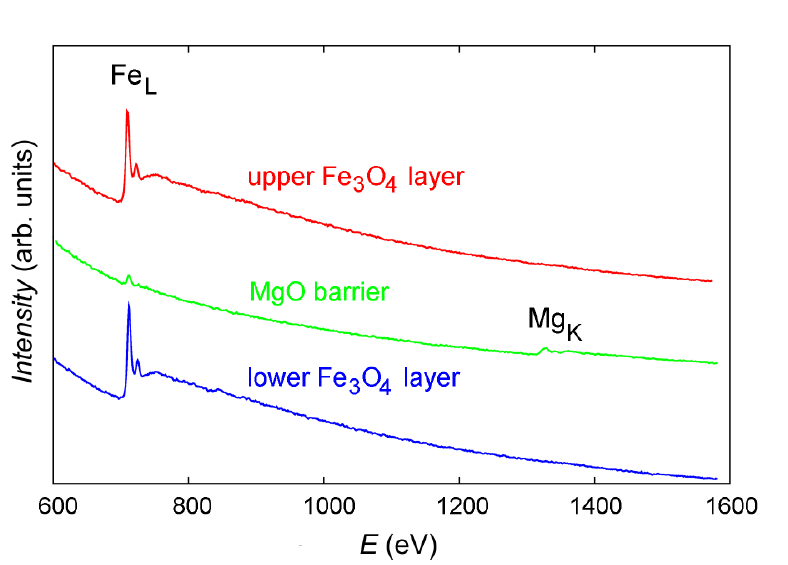}
    \caption{
EEL spectra of the MgO barrier and of the first and second Fe$_3$O$_4$ layer.
The spectrum of the MgO barrier contains a faint signal at the energy loss of
the Fe-$K$ edge (708\,eV) proving a small amount of iron.}
    \label{fig:MTJ-EELS}
\end{figure}

To obtain unambiguous information on the chemical composition of the MgO
barrier layer, the EEL spectrum in the center of the MgO barrier was analyzed
and is displayed in Fig.~\ref{fig:MTJ-EELS} together with spectra of the two
magnetite layers. In this spectrum a faint signal at the energy loss of the
Fe-$K$ edge (708\,eV) is present besides the Mg-$K$ edge proving a small amount
of iron in the MgO barrier. The obvious question arising at this point is
whether or not iron-doped MgO is a dilute magnetic semiconductor. If this is
the case, the observed ferromagnetic coupling across the MgO barrier is easy to
understand. To clarify this point, we have grown Fe-doped epitaxial MgO films
by PLD using an iron-containing MgO target. However, the magnetization of these
films was below the noise level of the SQUID magnetometer.

\subsection{Summary.}
 \label{sec:MTJ-Summary}

We have successfully fabricated Fe$_3$O$_4$-based MTJs with different size and
shape of the junction area. For MTJs with AlO$_x$ barriers a separated
switching of the magnetization of the two ferromagnetic electrodes is obtained.
These junctions show a reproducible TMR effect with an ideal symmetric
switching behavior. The TMR effect could be observed in the whole investigated
temperature range from 150\,K to 350\,K. At 350\,K, TMR values up to
27\% were obtained. A giant geometry-induced TMR effect can be generated
in Fe$_3$O$_4$-based MTJs due to the high resistivity of the electrode
material. Our data suggest that Fe$_3$O$_4$ with its high spin polarization and
high Curie temperature is interesting for spintronic devices. However, to
access the full potential of this material, the quality of suitable tunneling
barriers and involved interface layers has to be improved.

\section{Ferromagnetic Semiconductors.}
 \label{sec:ZnO}

\begin{figure}
 \centering
    \includegraphics[width=\linewidth]{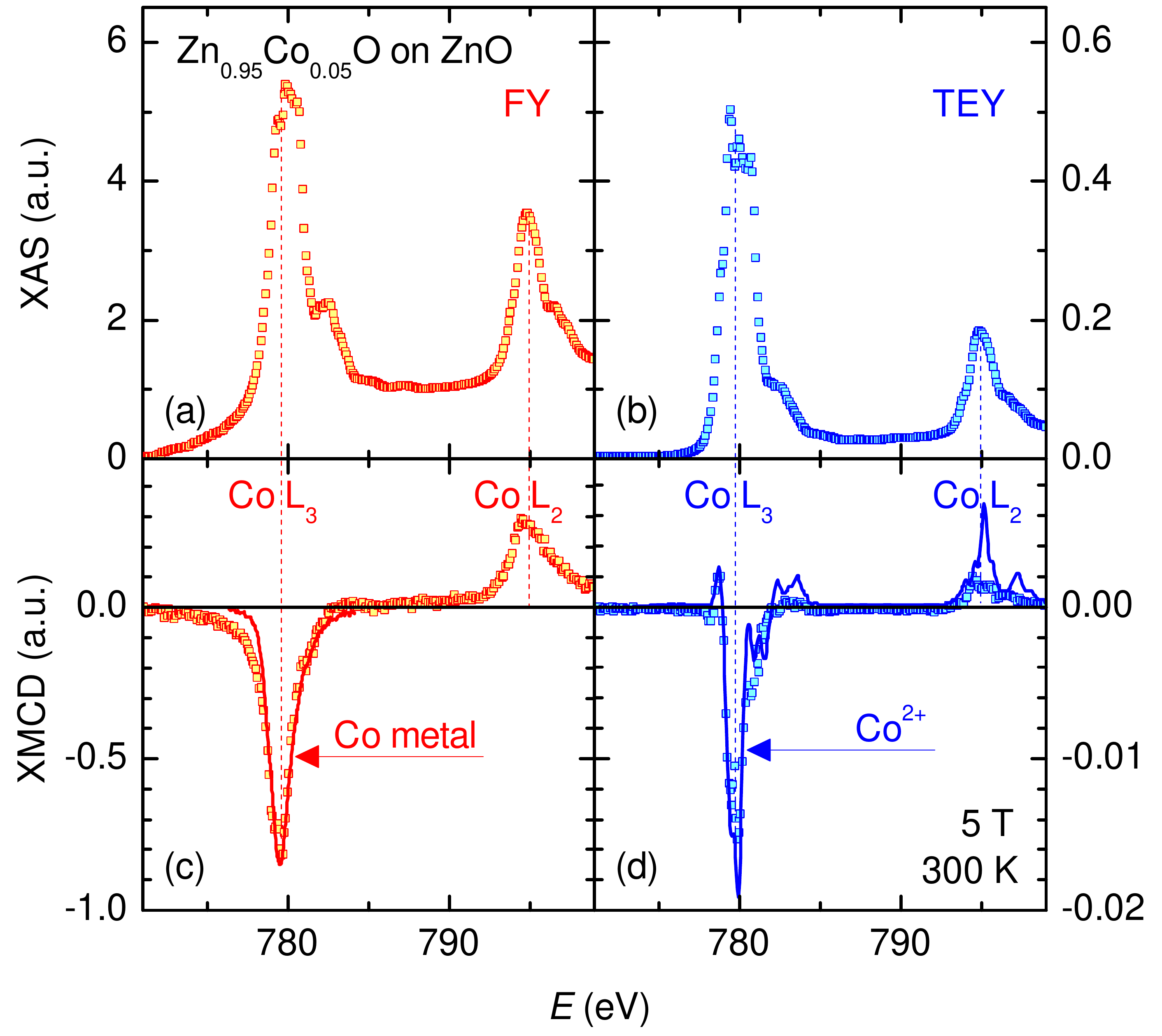}
    \caption{
    (a,b) XAS and (c,d) XMCD from Zn$_{0.95}$Co$_{0.05}$O at the Co $L_{2,3}$ edges,
    recorded simultaneously in fluorescence yield (FY) and total electron yield (TEY) mode at 300\,K
    and 5\,T. The symbols represent data points, the red line in (c) is an experimental spectrum for
    metallic Co taken from \cite{Mamiya2006}, the blue line in (d) is a calculated spectrum for Co$^{2+}$
    taken from \cite{Kobayashi2005}.}
    \label{fig:ZnO-XAS-XMCD}
\end{figure}

The rapidly evolving field of spintronics requires material systems combining
ferromagnetism (FM) with the versatile electronic properties of semiconductors.
Therefore, the integration of semiconducting and magnetic properties in one and
the same material to realize magnetic semiconductors was attracting broad
attention. Dilute magnetic semiconductors (DMS) such as the widely studied
material (Ga,Mn)As are attractive in this regard. Unfortunately, this well
established DMS has a Curie temperature $T_{\rm C} < 170$\,K~\cite{Ohno1996},
preventing room temperature applications. In contrast, $T_{\rm C} > 300$\,K has
been predicted for the oxide semiconductor ZnO on partial substitution of Zn by
a magnetic transition metal such as Mn or Co~\cite{Dietl2000}. Indeed,
ferromagnetic behavior has been observed in Zn$_{0.95}$Co$_{0.05}$O thin films
at room temperature~\cite{Venkatesan2004}. Some authors even report on two
magnetic regimes depending on the charge carrier density~\cite{Behan2008}, or
on gate-control of magnetism~\cite{Lee2009}. However, there is an ongoing
debate on the nature of the ferromagnetic behavior in oxide
semiconductors~\cite{Opel2008,Chambers2006,Ney2010} as the observed behavior
might also be traced back to nanometer-sized precipitates of the Co dopant
atoms embedded in a nonmagnetic ZnO matrix. The same controversy holds for Mn
in Ge~\cite{Ahlers2006,Jaeger2006}.

\begin{figure}
 \centering
    \includegraphics[width=\linewidth]{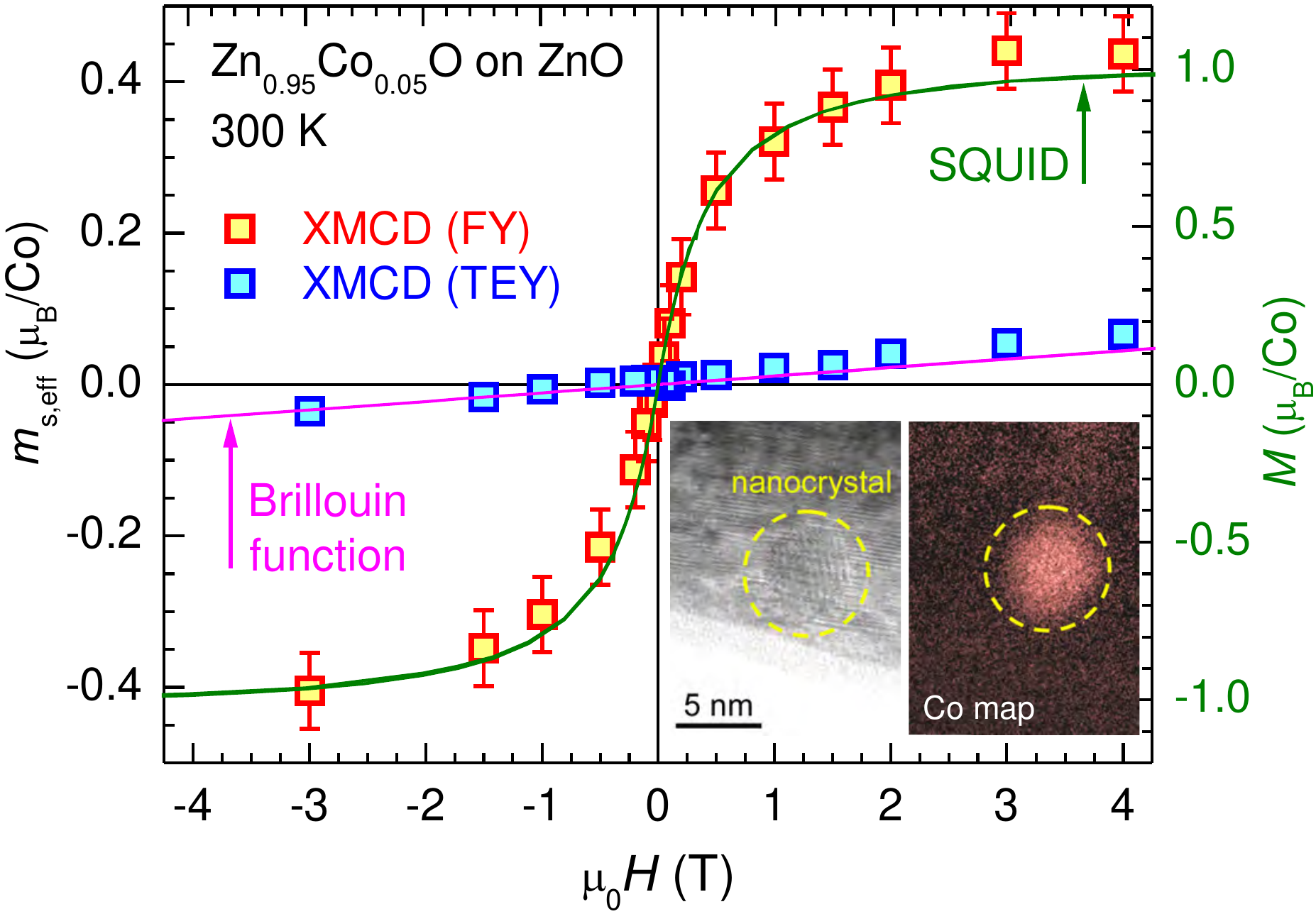}
    \caption{
Effective spin magnetic moment $m_{\rm s,eff}$ of Co in
Zn$_{0.95}$Co$_{0.05}$O, derived by X-ray magnetic circular dichroism, recorded
in total electron yield (TEY, blue) and fluorescence mode (FY, red). The FY
data follow the sample magnetization $M$ (green), determined by SQUID
magnetometry. The TEY data fit well to a Brillouin function for paramagnetic
Co$^{2+}$ ions (pink). The inset shows a TEM image (left) displaying a
nanocrystal in the ZnO matrix and an elemental map of Co obtained by EF-TEM
(right), displaying Co enrichment in the same area. The figure is reproduced
from \cite{Opel2008,Ney2010}.}
    \label{fig:ZnO-XMCD}
\end{figure}

To unambiguously clarify the nature of magnetism in Zn$_{0.95}$Co$_{0.05}$O, we
prepared thin films using laser-MBE and carefully investigated their structural
and magnetic properties as discussed in detail in \cite{Opel2008}.
In particular, we have combined SQUID magnetometry,
X-ray magnetic circular dichroism (XMCD), and AC susceptibility measurements
with careful X-ray and high resolution TEM studies. We simultaneously recorded
XMCD spectra in both the total electron (TEY) and the fluorescence yield (FY)
modes, allowing for an element-specific distinction between surface and bulk
magnetic properties, respectively (Fig.~\ref{fig:ZnO-XAS-XMCD}). Our data provide clear
evidence that our Zn$_{0.95}$Co$_{0.05}$O thin films are not homogeneous DMS.
The large magnetic moments observed at room temperature in some of the samples
could rather be traced back to the presence of nanometer-sized
superparamagnetic Co precipitates within the bulk of Zn$_{0.95}$Co$_{0.05}$O,
which were directly evidenced by XMCD and energy-filtering transmission
electron microscopy (EF-TEM) (Fig.~\ref{fig:ZnO-XMCD}). Other samples show
pure paramagnetism of isolated Co$^{2+}$ moments from room temperature down to
5\,K. More details have been published elsewhere~\cite{Opel2008,Ney2010}.
Similar behavior has been reported for other dilute magnetic systems, like
GaN:Gd~\cite{Ney2008b}. Of course, our data do not prove that the realization
of a DMS is impossible for ZnO:Co. However, more effort is required to
unambiguously determine the nature of ferromagnetism, and conclusions based on
superficial studies only presenting $M(H)$ ``hysteresis'' curves should be
considered with care.

Since the room temperature realization of ferromagnetic and semiconducting
properties in one material may turn out difficult, the use of
ferromagnet/semiconductor heterostructures may be the appropriate solution.
Here, the use of oxide materials may be advantageous. We recently have shown
that semiconducting ZnO can be grown epitaxially on ferromagnetic Fe$_3$O$_4$
and vice versa~\cite{Nielsen2008}. Such oxide heterostructures are interesting
since the semiconducting and ferromagnetic oxide have similar resistivity,
avoiding the well-known resistivity mismatch problem at ferromagnetic
metal/semiconductor interfaces \cite{Schmidt2000} regarding efficient spin injection. Future
studies have to show whether the interfaces in the oxide heterostructures can
be controlled and engineered in an adequate way to allow for the realization of
spin injection devices.

\section{Multiferroics and Strain Effects.}
 \label{sec:MF}

\begin{figure}
 \sidecaption
    \includegraphics[width=0.5\linewidth]{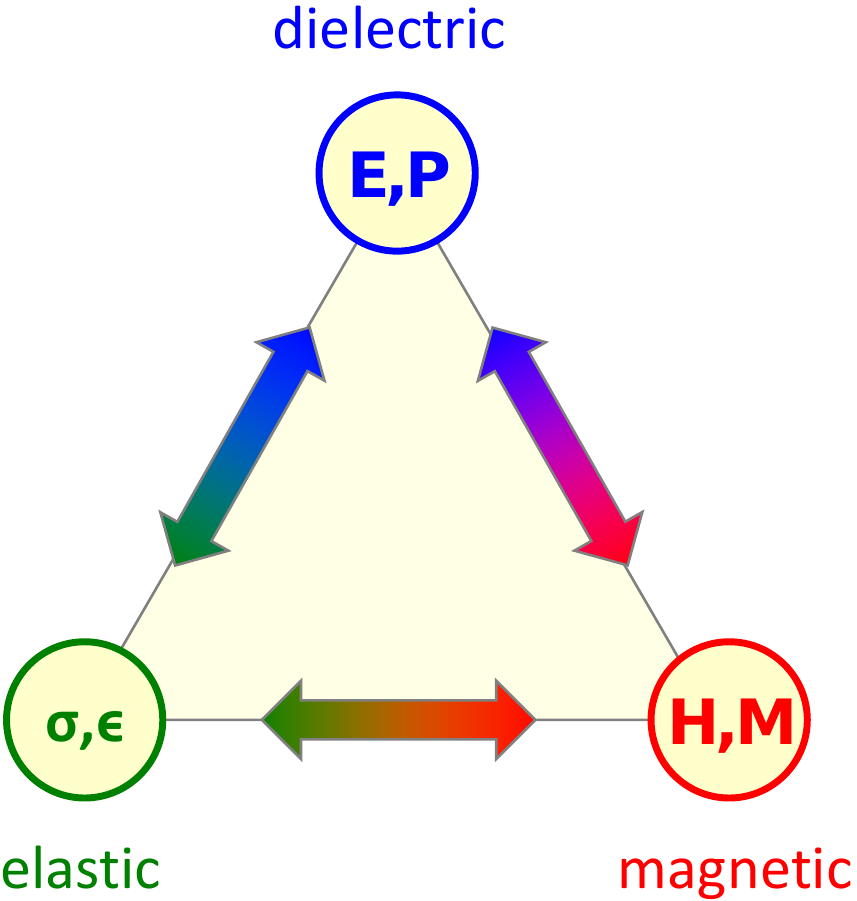}
    \caption{
The response of condensed matter to external electric fields ($\vec{E}$),
magnetic fields ($\vec{H}$), or mechanical stress ($\vec{\sigma}$) are
polarization ($\vec{P}$), magnetization ($\vec{M}$), or strain
($\vec{\epsilon}$), respectively. For useful multiferroic materials exhibiting spontaneous
$\vec{P}$, $\vec{M}$, or $\vec{\epsilon}$ also a strong coupling between the
dielectric, magnetic, and elastic properties is required~\cite{Spaldin2005}.}
    \label{fig:MF-Dreieck}
\end{figure}

An active and promising field of research is the integration of different
ferroic ordering phenomena such as ferromagnetism, ferroelectricity, or
ferroelasticity in one and the same
material~\cite{Fiebig2005,Eerenstein2006,Ramesh2007}. The coexistence of
ferroelectricity and ferromagnetism in novel multi-functional materials is
particularly interesting, since this could allow the realization of new
functionalities of electro-magnetic devices, such as the electric field-control
of magnetization. Unfortunately, it turned out
that there are very few intrinsic ferroelectric/ferromagnetic multiferroics
because the standard microscopic mechanisms driving ferroelectricity and
ferromagnetism are incompatible. They usually require empty and partially
filled transition metal orbitals, respectively~\cite{Hill2000}. Furthermore,
most of the multiferroics reported so far are antiferromagnets,
which are not expected to respond noticeably when
applying magnetic fields. This has initiated the search for other mechanisms
favoring the coexistence of ferroelectricity and magnetic order as, for
example, in BiMnO$_3$~\cite{Kimura2003} or in extrinsic, two-component
multiferroic thin film heterostructures~\cite{Spaldin2005}. For the realization
of useful multiferroic materials, however, the different ferroic order
parameters do not only have to coexist, but must be coupled to each other
(Fig.~\ref{fig:MF-Dreieck}) to enable, e.g. an electric field control of the
magnetization in magnetoelectric multiferroics.

Here, we report on different oxide systems which are promising in this regard.
We first study the dielectric and magnetic properties of the perovskites
BiFeO$_3$ and BiCrO$_3$ which are considered as intrinsic multiferroic
materials and search for a magnetoelectric coupling in these materials. Second,
we focus on hybrid systems consisting of ferroelectric and ferromagnetic
materials. As an example, we discuss the behavior of multiferroic hybrids consisting
of epitaxial thin films of the ferromagnet Sr$_2$CrReO$_6$ grown on ferroelectric
BaTiO$_3$ substrates.

\subsection{Intrinsic Multiferroics: BiFeO$_3$ and BiCrO$_3$.}
 \label{sec:MF-BFO-BCO}

The perovskite BiFeO$_3$ is one of the few robust materials with ferroelectric
and antiferromagnetic order well above room temperature. In bulk material,
BiFeO$_3$ is antiferromagnetic below the N\'{e}el temperature of $T_{\rm N} =
643$\,K and ferroelectric below $T_{\rm C} =
1103$\,K~\cite{Kiselev1963,Smolenskii1961}. Thin films of BiFeO$_3$ are
discussed for application as exchange biasing ferromagnetic layers in
spintronic multilayer structures~\cite{Bea2006,Bea2008,Bibes2008,Chu2008}.
Strained thin films of BiFeO$_3$ have attracted renewed interest after the
report of a high ferroelectric polarization of $60\,{\rm \mu C/cm^2}$ together
with a high residual magnetic moment of $M_{\rm S} = 1\,{\rm \mu_B}$ per formula
unit (f.u.)~\cite{Wang2003}. However, there is an ongoing controversial
discussion about the origin of this saturation
magnetization~\cite{Eerenstein2005,Wang2005} as the initially reported
magnetization values in relaxed BiFeO$_3$ thin films could not be reproduced by
subsequent work~\cite{Gepraegs2007,Eerenstein2005,Bea2005,Bea2009}. In our work
we found $M_{\rm S} = 0.02 \,{\rm \mu_B}$/f.u. in strained BiFeO$_3$ thin
films~\cite{Gepraegs2007} which is in full agreement with other recent
publications~\cite{Eerenstein2005,Bea2005,Bea2009} and density functional
calculations~\cite{Ederer2005a}. Moreover, it was shown that strain-free bulk
crystals of BiFeO$_3$ do not show any parasitic ferromagnetism down to 2\,K at
all~\cite{Lu2009}. Whether or not the small magnetic moments reported for
BiFeO$_3$ thin films and also observed in our samples originate from nanoscale
metallic Fe precipitates or spin canting has to be clarified.
Details on the growth as well as the structural
and magnetic characterization of our BiFeO$_3$ thin films can be found
in~\cite{Gepraegs2007}.

Another interesting candidate for an intrinsic multiferroic material is
BiCrO$_3$. Bulk BiCrO$_3$ is antiferromagnetic below $T_{\rm N} =
123$\,K~\cite{Sugawara1968} and shows a weak ferromagnetic moment, which is
attributed to Cr$^{3+}$ spin canting~\cite{Gepraegs2007}. The situation is
similar in thin films, where a N\'{e}el temperature between
120\,K~\cite{Murakami2006} and 140\,K~\cite{Kim2006} is reported. Regarding
dielectric properties, the picture is still unclear as both ferroelectricity at
room temperature~\cite{Murakami2006} and antiferroelectricity at
5\,K~\cite{Kim2006} have been observed. In the following we present results on
the fabrication and characterization of this material.

\paragraph{Thin Film Deposition.}

We have grown $c$-axis oriented epitaxial BiCrO$_3$ thin films by laser-MBE
with \textit{in-situ} RHEED~\cite{Gross2000,Klein1999} on (001)-oriented
SrTiO$_3$ substrates. The PLD process was performed in the same way as
described in subsection~\ref{sec:MTJ-PLD} for Fe$_3$O$_4$. We used
polycrystalline targets with excess bismuth in the composition
Bi$_{1+\delta}$CrO$_3$ ($\delta = 15 \ldots 20\%$). The growth parameters were
optimized with regard to the structural quality of the samples. The thin films
were grown by imposed layer-by-layer interval deposition in O$_2$ atmosphere at
a pressure of $8\times10^{-3}$\,mbar and a substrate temperature of
$570^\circ$C with a laser fluence of 2\,J/cm$^2$ at the
target~\cite{Gepraegs2007}. To enable a dielectric characterization of the samples in
spite of the insulating SrTiO$_3$ substrate, some thin films have been
deposited with an underlying bottom electrode layer of the conducting oxide
SrRuO$_3$ ($\rho = 2.75\,{\rm \mu \Omega m}$ at 300\,K) with a thickness
between 10\,nm and 20\,nm. The SrRuO$_3$ film was also grown by PLD from a
stoichiometric target in O$_2$ atmosphere (0.15\,mbar) at a substrate
temperature of $700^\circ$C.

\paragraph{Structural Characterization.}

\begin{figure}
    \centering
    \includegraphics[width=\linewidth]{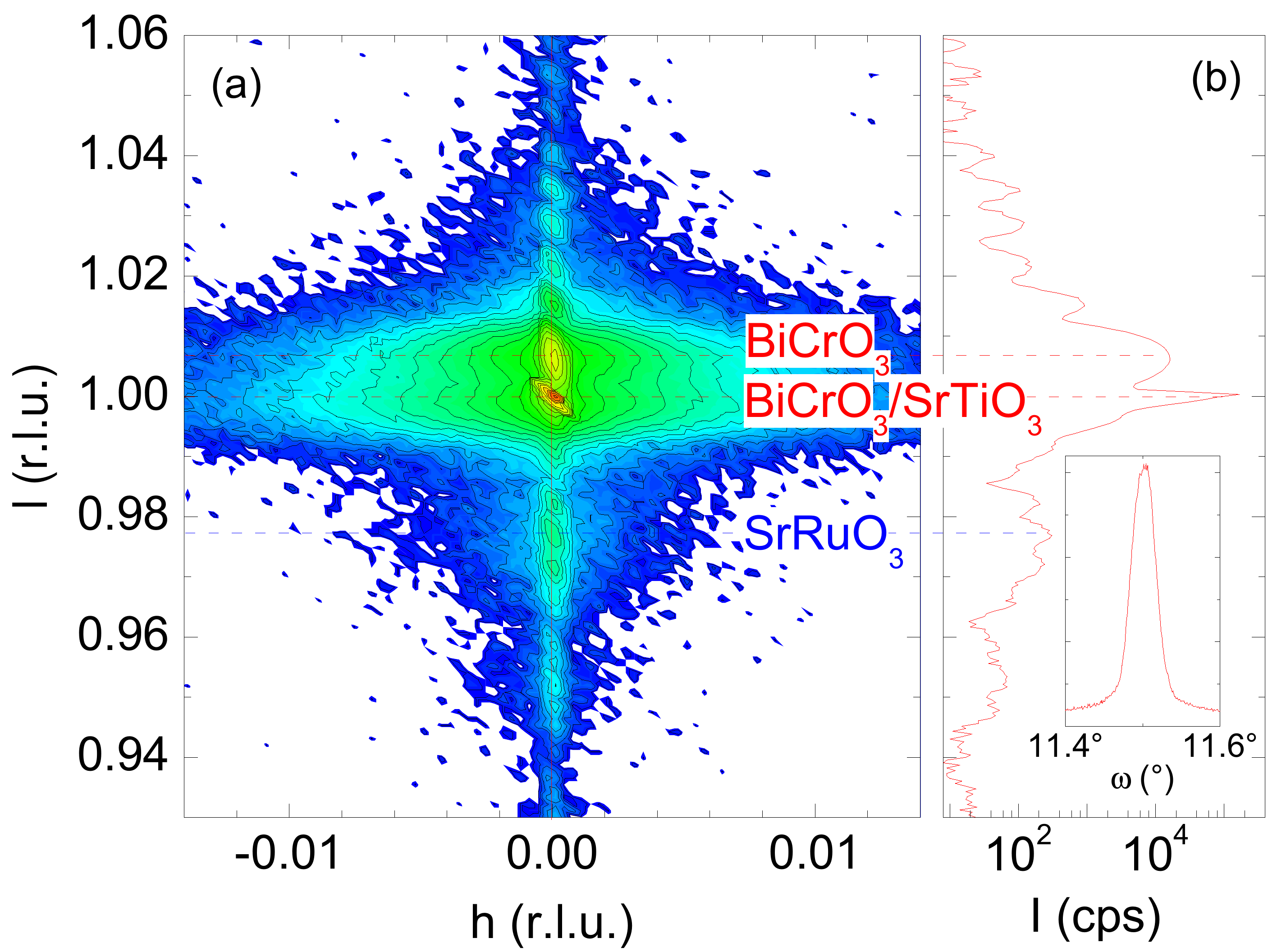}
    \caption{
(a) Reciprocal space map of a BiCrO$_3$/SrRuO$_3$ bilayer (thickness: 65\,nm/19\,nm)
grown on SrTiO$_3$ around the (001) reflection of the substrate. (b) Line scan
for $h=0$. The inset shows a rocking curve of the second BiCrO$_3$(001)
reflection with $\ell>1$.}
    \label{fig:MF-BiCrO3-RSM}
\end{figure}

The structural properties of the films were investigated by X-ray diffraction
using Cu K$\alpha_1$ radiation and a high-resolution four-circle diffractometer
(Bruker AXS discover). Fig.~\ref{fig:MF-BiCrO3-RSM}(a) shows a reciprocal space
map around the (001) reflection of the SrTiO$_3$ substrate obtained at room
temperature. A line scan for $h=0$ ($00\ell$-scan) is shown in
Fig.~\ref{fig:MF-BiCrO3-RSM}(b). The film thicknesses are 19\,nm (SrRuO$_3$)
and 65\,nm (BiCrO$_3$). BiCrO$_3$ shows two reflections at $\ell=1$ and
$\ell=1.007$, as indicated by the broad green regions in (a). The reflection
from SrRuO$_3$ is located at $\ell=0.978$. Satellites due to Laue oscillations
from the BiCrO$_3$ as well as the SrRuO$_3$ reflections are visible,
demonstrating that the films are coherently strained and have small surface and
interface roughnesses. Moreover, the rocking curve of the BiCrO$_3(001)$
reflection displays a small full width at half maximum of $0.032^\circ$ [inset
in Fig.~\ref{fig:MF-BiCrO3-RSM}(b)] demonstrating a very low mosaic spread.

\begin{figure}
    \centering
    \includegraphics[width=\linewidth]{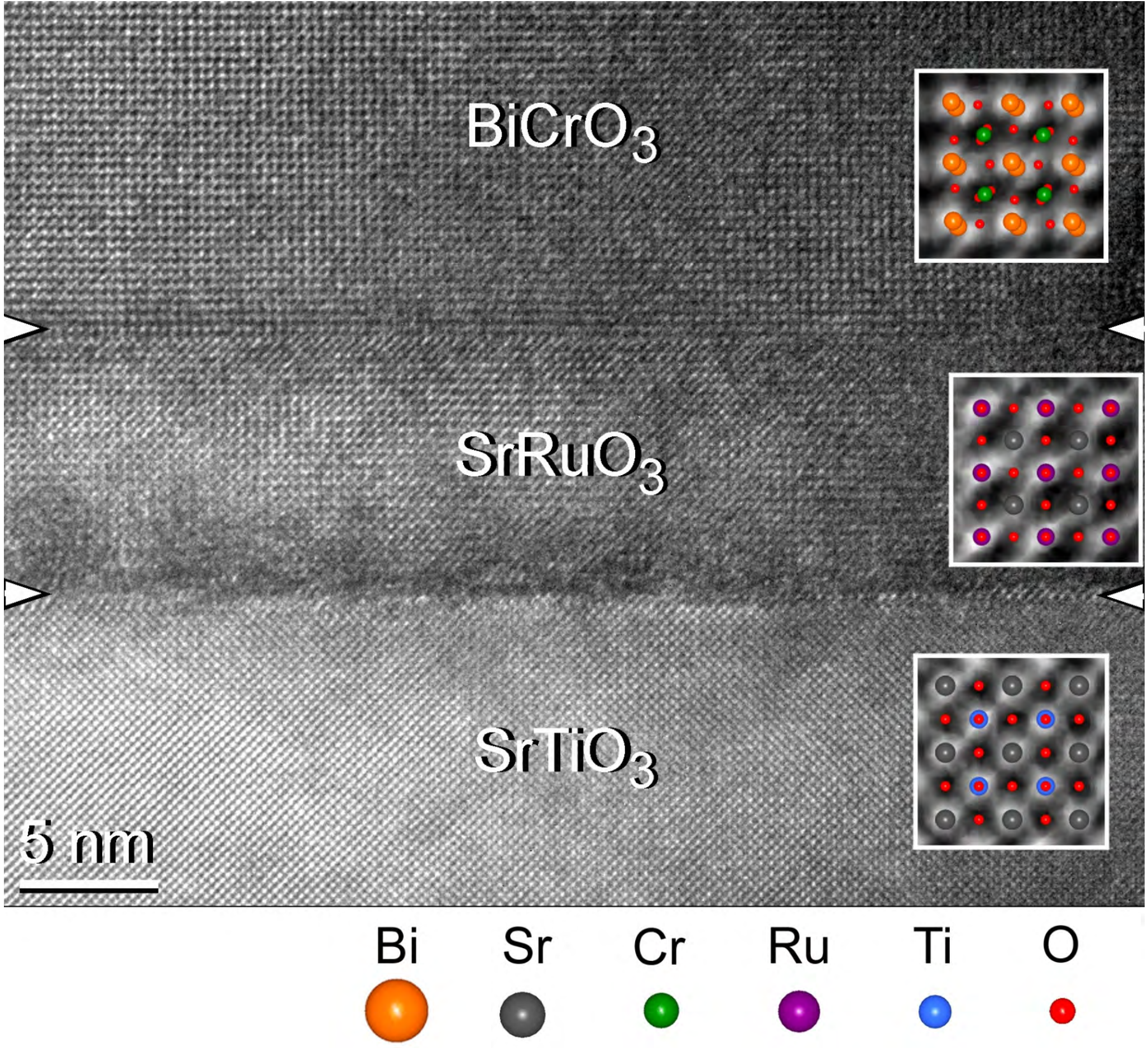}
    \caption{
HR-TEM image of a heteroepitaxial BiCrO$_3$/SrRuO$_3$ (28\,nm/10\,nm) bilayer grown
on a SrTiO$_3$(100) substrate; all crystals in [001] direction, interfaces are
marked. Magnified lattice images of the crystals with inset of atomic columns
prove perfect epitaxy of the three perovskite oxides and continuation of the
oxygen sub-lattices across the interfaces. }
    \label{fig:MF-BiCrO3-TEM}
\end{figure}

The high structural quality of the BiCrO$_3$ layer as well as of the SrRuO$_3$
electrode layer is demonstrated in real space at high spatial resolution by TEM
studies (Fig.~\ref{fig:MF-BiCrO3-TEM}). First the thicknesses of the two layers
were measured from TEM images to 10\,nm (SrRuO$_3$) and 28\,nm (BiCrO$_3$) in this sample. The
crystals containing the interfaces SrTiO$_3$/SrRuO$_3$ and BiCrO$_3$/SrRuO$_3$
are imaged in the HR-TEM micrograph (Fig.~\ref{fig:MF-BiCrO3-TEM}). It is worth
noting that the image contrast in the three crystals, all with $AB$O$_3$
perovskite structure, is different. This is the result of the metal ions $A$
and $B$ which have very different scattering power in the three crystals, and
the scattering potential scales with the atomic number $Z$. In SrTiO$_3$, the
scattering powers of Sr ($Z_{\rm Sr} = 38$) and of the Ti+O column ($Z_{\rm Ti}
+ Z_{\rm O} = 22 + 8 = 30$) are similar, and therefore the $\{110\}$ type of
lattice planes dominate the crystal lattice contrast, forming a square pattern
$45^\circ$ rotated to the $a$ axis. In SrRuO$_3$, the $B$ site atom ruthenium
has the highest scattering potential and the atomic columns Ru+O ($Z_{\rm Ru} +
Z_{\rm O} = 44 + 8 = 52$) dominate the contrast and produce a square array of
$\{100\}$ lattice planes centered at the $B$ sites. In BiCrO$_3$, the heavy
metal bismuth on the $A$ site is by far the strongest scatterer ($Z_{\rm Bi} =
83$) and produces again a square array contrast of $\{100\}$ lattice planes,
here, however, centered at the $A$ sites. Close inspection of the lattice plane
image (Fig.~\ref{fig:MF-BiCrO3-TEM}) reveals indeed a shift of the dot contrast
from SrRuO$_3$ to BiCrO$_3$ of half the $a$ axis parallel to the interface. To
better illustrate the contrast in the perovskite crystals and the position of
the atomic columns, insets at high magnification are placed into
Fig.~\ref{fig:MF-BiCrO3-TEM}. These observations prove continuation of the
oxygen sub-lattices of the three perovskite crystals across the interfaces. A
further detail concerning the interfaces is worth noting: the interface
SrTiO$_3$/SrRuO$_3$ appears not to be atomically abrupt. This is very likely
the result of surface treatment of the SrTiO$_3$(001) crystal by grinding and
polishing. Nevertheless, the SrRuO$_3$ electrode has grown epitaxial, and even
more impressive is the interface between the BiCrO$_3$ and SrRuO$_3$ layers
which appears atomically planar, i.e. without steps, and with a direct
structural transition from one phase to the other. In summary, the observations
and structural characterization by X-ray diffraction and TEM prove that the
thin film system of BiCrO$_3$ and SrRuO$_3$ is state-of-the-art grown with
outstanding quality of the crystals.

\paragraph{Magnetic Characterization.}

\begin{figure}
    \centering
    \includegraphics[width=\linewidth]{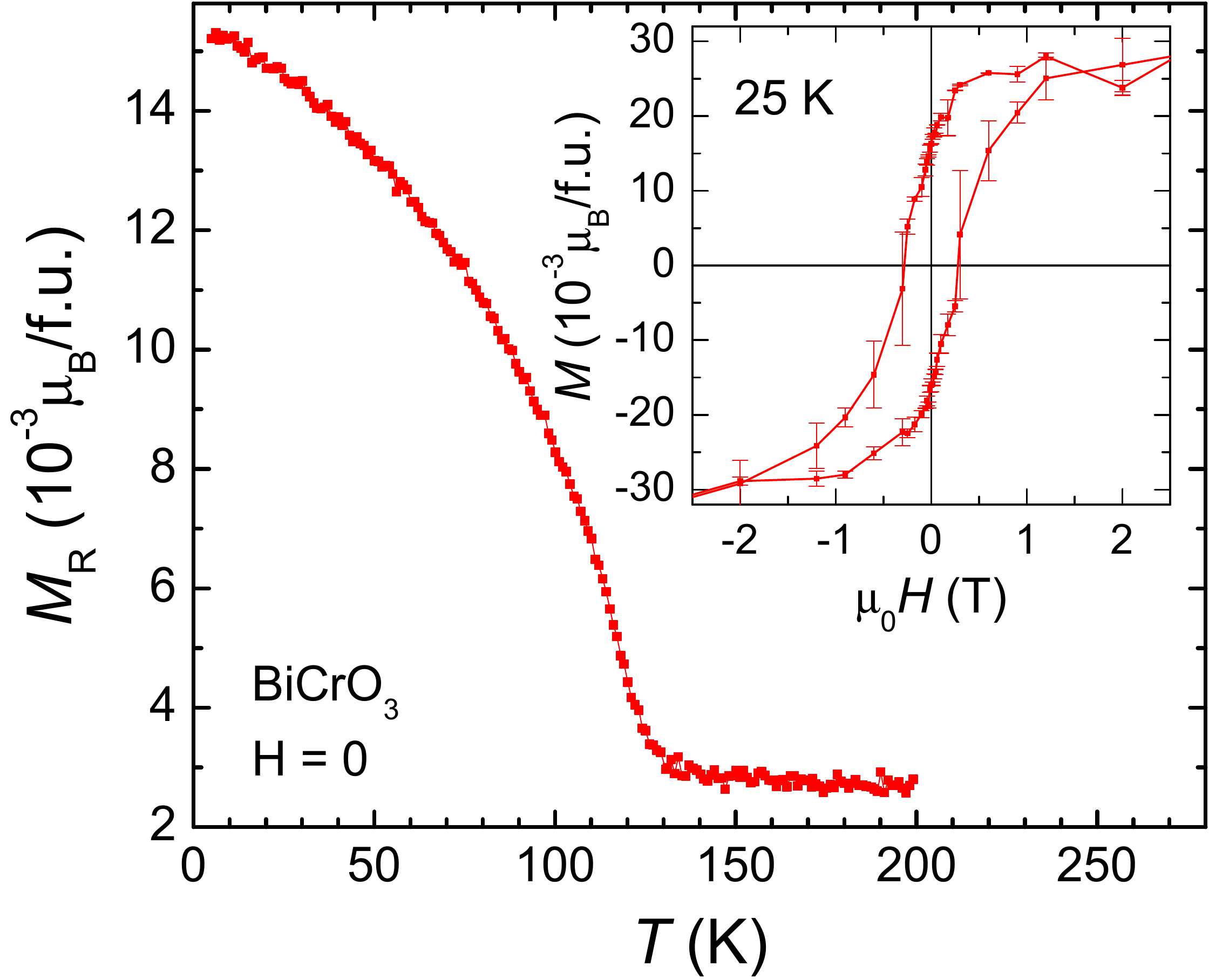}
    \caption{
Remanent magnetization $M_{\rm R}$ of a 59\,nm thick film of BiCrO$_3$ grown on
a SrTiO$_3$:Nb substrate as a function of temperature $T$. The magnetization was measured
after field cooling at 7\,T in zero field. The inset shows a $M(H)$ loop recorded at 25\,K.}
    \label{fig:MF-BiCrO3-SQUID}
\end{figure}

Magnetization measurements were performed with a SQUID magnetometer in magnetic
fields up to $\mu_0 H = 7$\,T applied in the film plane.
Fig.~\ref{fig:MF-BiCrO3-SQUID} displays the remanent magnetization $M_{\rm
R}(T)$, measured after field cooling at 7\,T on increasing temperature in zero
field. Although antiferromagnetic, the sample shows a weak residual remanent
magnetization of $0.015\,\mu_{\rm B}$/f.u. at low temperatures. We note that
this observed parasitic ferromagnetism sets in below a critical temperature of
about $T_{\rm C} = 128$\,K. This value is slightly larger than the N\'{e}el
temperature of bulk material ($T_{\rm N} = 123$\,K~\cite{Sugawara1968}) and
than the values for parasitic ferromagnetic ordering reported in
literature~\cite{Murakami2006,Niitaka2004}. This may be related to the small
tensile strain in our BiCrO$_3$ films. A magnetic field-loop recorded at 25\,K demonstrates that $M$
saturates at about $M_{\rm S} = 0.03\,\mu_{\rm B}$/f.u. for fields higher than
2\,T (see inset of Fig.~\ref{fig:MF-BiCrO3-SQUID}). The coercivity of this
parasitic ferromagnetism is larger than for our BiFeO$_3$
films~\cite{Gepraegs2007}. The scatter seen in the data is caused by the small
sample volume leading to absolute values of $M$ close to the resolution limit
of the SQUID magnetometer ($\sim 10^{-11}$\,Am$^2$). In summary, the observed
magnetic properties of our BiCrO$_3$ films are consistent with the picture that
the Cr$^{3+}$ spins are coupled antiferromagnetically and that a slight canting
of the spins results in a weak ferromagnetic
signal~\cite{Ederer2005a,Ederer2005b}.

\paragraph{Dielectric Characterization.}

\begin{figure}
    \centering
    \includegraphics[width=\linewidth]{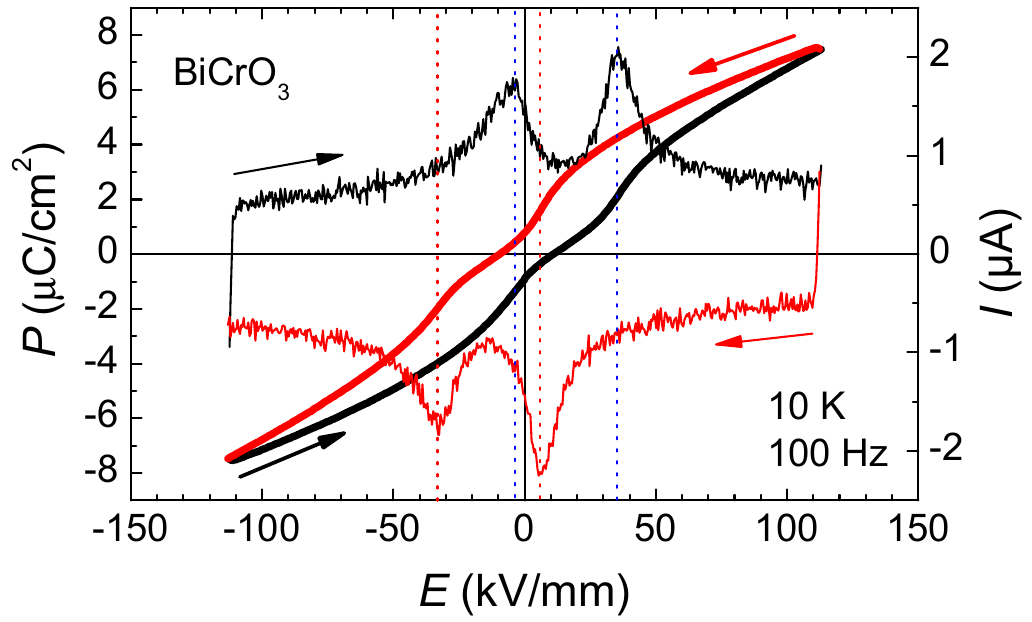}
    \caption{
Dielectric polarization (thin solid lines, left scale) and dielectric current
(thick solid lines, right scale) as a function of the electric field of a BiCrO$_3$ thin
film. The data are recorded in dynamic mode at 100\,Hz and 10\,K. The field-up
sweeps are shown in black, the field-down sweeps in red. The dashed vertical lines
mark the field position where the polarizations of the dielectric sub-lattices change their sign.}
    \label{fig:MF-BiCrO3-AFE}
\end{figure}

The dielectric properties of the thin film samples were investigated using a
commercial ferroelectric tester (aixACCT TF analyzer 2000).
Figure~\ref{fig:MF-BiCrO3-AFE} displays the electric $P(E)$ loop from a 177\,nm
thin film of BiCrO$_3$ together with the dielectric current, recorded in
dynamic mode at 100\,Hz at a temperature of 10\,K in electric fields of up to
$E = 115$\,kV/mm (corresponding to a voltage of 25\,V) applied across the
BiCrO$_3$ layer. The $P(E)$ curve shows an antiferroelectric hysteresis loop
with an electric polarization close to $8\,{\rm \mu C/cm^2}$ at high electric
fields. Two
``humps'' per field sweep direction at $E=5$\,kV/mm and $E=35$\,kV/mm are
visible in the $P(E)$ loop. These observations clearly point to the presence of
anti-ferroelectric ordering in this material which
becomes more obvious when investigating the dielectric current $I$. This
quantity shows two clear and well pronounced maxima per field sweep direction
at the same $E$ values where the ``humps'' in $P(E)$ are observed. We interpret
this double peak feature as evidence for the switching of two ferroelectric
sub-polarizations in a dipole lattice.

\paragraph{Summary.}

Neither BiFeO$_3$ nor BiCrO$_3$ shows the expected coexistence of
ferromagnetism and ferroelectricity. (i)~BiFeO$_3$ is antiferromagnetic. The
existence of the reported parasitic ferromagnetic phase with moments up to
$1\,\mu_{\rm B}$/f.u.~\cite{Wang2003} could not be reproduced and most likely
is an artefact caused by metallic Fe precipitates in the BiFeO$_3$ thin films.
(ii)~BiCrO$_3$ is antiferroelectric on large scales instead of showing
ferroelectric behavior. (iii)~Neither for BiFeO$_3$ nor BiCrO$_3$, we were able
to find any indication for intrinsic magnetoelectric coupling.

\subsection{Extrinsic Multiferroics: Piezo-strain in Fe$_3$O$_4$.}
 \label{sec:MF-Fe3O4-Piezo}

Another promising approach to realize multiferroic coupling is the fabrication
of heterostructures composed of materials with different ferroic properties. Of
course, in such multilayers coexisting ferroic ordering phenomena can be easily
realized. The key task here is to establish a coupling of the order parameters
at the interfaces between the ferroic sublayers. Along this line, the use of
strain to establish a finite coupling via the elastic channel (see
Fig.~\ref{fig:MF-Dreieck}) is promising. We showed in
\cite{Goennenwein2008,Brandlmaier2008} that the magnetic
anisotropy and in turn the magnetization direction of a ferromagnetic layer can
be varied by an electric field making use of the elastic channel.
In our experiments we used magnetite (Fe$_3$O$_4$) as a prototype ferromagnet.
Epitaxial Fe$_3$O$_4$ was deposited by laser-MBE with \textit{in-situ} RHEED
\cite{Gross2000,Klein1999} on (001)-oriented MgO substrates as described
in~\ref{sec:MTJ-PLD}. The thin films with a thickness of 44\,nm were grown in
an inert Ar atmosphere at a pressure of $0.06$\,mbar with a laser fluence of
2.5\,J/cm$^2$ at the target and a substrate temperature of $320^\circ$C. After
deposition, the MgO substrate was polished down to a thickness of about
$50\,{\rm \mu m}$. To introduce an \textit{in-situ} tunable strain, we cement a
commercial Pb(Zr,Ti)O$_3$ piezoelectric actuator on top of the Fe$_3$O$_4$
thin film~\cite{Shayegan2003,Botters2006}. The expansion (or contraction) of
the piezoelectric actuator as a function of the applied electric voltage
$V_{\rm piezo}$ is directly transferred into the magnetite film, yielding a
voltage-controllable strain contribution. To optimize the induced strain, we
cement the piezoactuator face to face onto the Fe$_3$O$_4$ thin film. The
induced strain was determined by X-ray diffraction via the changes of the
lattice parameters of the underlying MgO substrate~\cite{Brandlmaier2008}.

\begin{figure}
 \centering
    \includegraphics[width=\linewidth]{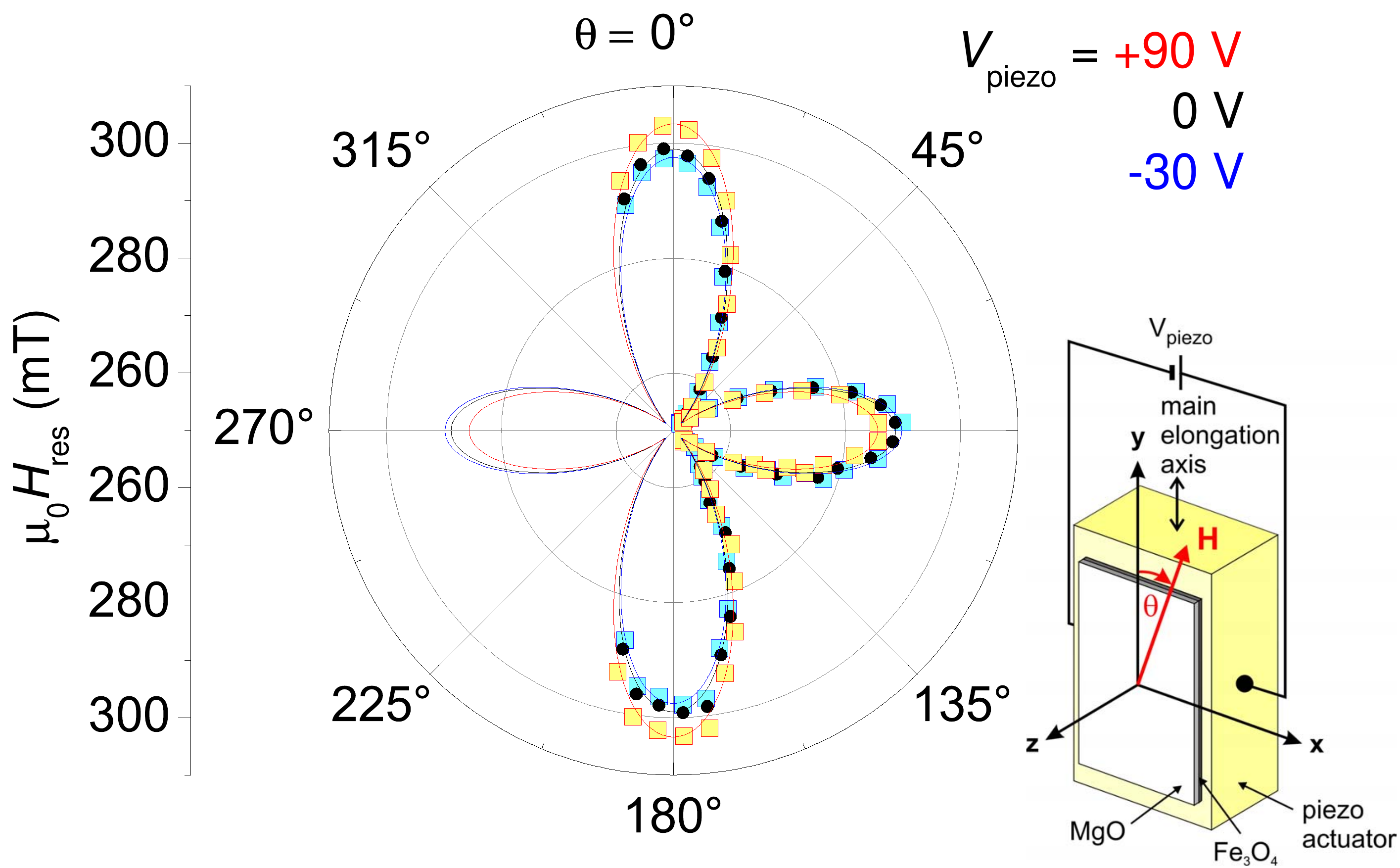}
    \caption{
    Ferromagnetic resonance (FMR) fields of the Fe$_3$O$_4$/piezo system
    at room temperature as a function of the angle $\theta$ between magnetic field $H$
    and the main elongation axis of the piezo actuator. The symbols represent data points
    obtained at different applied piezo voltages. The lines are simulations based on
    magnetoelastic theory.}
    \label{fig:MF-Fe3O4-piezo}
\end{figure}

To determine the magnetic anisotropy, we performed ferromagnetic resonance (FMR)
spectroscopy in the X-band (microwave frequency of 9.3~GHz) at room temperature.
The angular dependence of the resonance field $\mu_0 H_{\rm res}$ for different
values for $V_{\rm piezo}$ was successfully modeled using magnetoelastic
theory (Fig.~\ref{fig:MF-Fe3O4-piezo}) \cite{Brandlmaier2008}.
Implying a strain transmission efficiency factor of 70\%
from the piezoactuator to the Fe$_3$O$_4$ thin film, we find excellent
agreement with experiment. The changes of magnetic anisotropy induced by this
piezo-strain make it possible to shift the minimum of the free energy and thus
the magnetization orientation by about $6^\circ$ within the Fe$_3$O$_4$ film
plane. As the
magnetostriction constant $\lambda_{100}^{\rm Fe_3O_4} = -19.5 \times 10^{-6}$
\cite{Bickford1955} is comparable to those of e.g. Fe, Ni, or CrO$_2$
\cite{Lee1955,Miao2005}, our results show that a piezo-strain control of the
magnetization orientation is a realistic and versatile scheme applicable to a
variety of ferromagnets~\cite{Goennenwein2008,Brandlmaier2008,Weiler2009}.

\subsection{Extrinsic Multiferroics: Sr$_2$CrReO$_6$ on BaTiO$_3$.}
 \label{sec:MF-Sr2CrReO6-BTO}

\begin{figure}
 \centering
    \includegraphics[width=\linewidth]{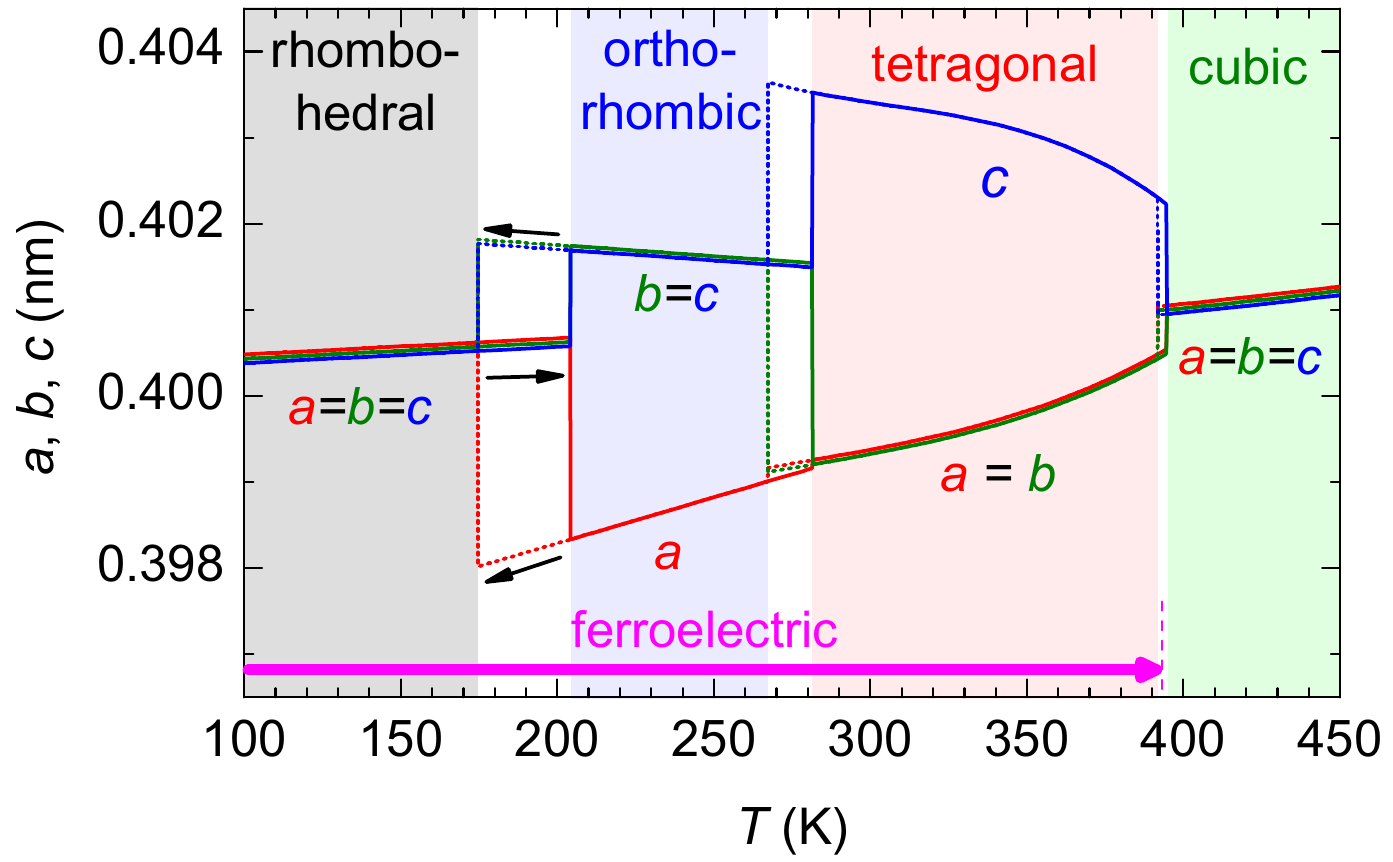}
    \caption{
Phase diagram of BaTiO$_3$ according to \cite{Shebanov1981}. The ferroelectric
Curie temperature is 393\,K.}
    \label{fig:MF-BaTiO3}
\end{figure}

Even higher strains are possible when replacing the piezoelectric
actuator by a ferroelectric substrate. This integration of ferromagnetic (FM) and
ferroelectric (FE) degrees of freedom in epitaxial hybrid heterostructures is
promising for the realization of multi-functional material systems with
improved or novel functionality.
As an example, we exploited the strain-induced change of the magnetic
coercivity of polycrystalline Ni in a FM/FE hybrid and demonstrated
both reversible and irreversible magnetization control \cite{Gepraegs2010}.
Here, we will focus on an epitaxial all-oxide FM/FE system.

Barium titanate (BaTiO$_3$) is the prototype ferroelectric. Its
crystallographic structure shows a variety of phase transitions, dependent on
the temperature (see Fig.~\ref{fig:MF-BaTiO3}) \cite{Shebanov1981}. Above
393\,K, bulk BaTiO$_3$ is cubic and paraelectric. Below 393\,K, it becomes
ferroelectric and its lattice structure changes to tetragonal. Within the
ferroelectric state, the lattice symmetry is further reduced to orthorhombic
(below 278\,K), and finally to rhombohedral (below 183\,K). The dielectric
constant, the spontaneous polarization, as well as the lattice constants change
abruptly at these phase transition temperatures, accompanied by a thermal
hysteresis~\cite{Kay1949}. Recently, there have been reports on
CoFe$_2$O$_4$/BaTiO$_3$, SrRuO$_3$/BaTiO$_3$,
La$_{0.67}$Sr$_{0.33}$MnO$_3$/BaTiO$_3$, and Fe$_3$O$_4$/BaTiO$_3$
hybrids~\cite{Chopdekar2006,Lee2000,Dale2003,Eerenstein2007,Tian2008,Vaz2009}.
They show a magnetoelectric coupling effect, manifesting itself as a change in
the magnetization and resistance of the ferromagnetic thin film at the
structural phase transition temperatures of BaTiO$_3$. In this regard, another
promising material is the ferromagnetic double perovskite Sr$_2$CrReO$_6$ which
is known for a giant anisotropic magnetostriction~\cite{Serrate2007} caused by
a large orbital moment on the Re site~\cite{Majewski2005c}. In addition, it has
a high Curie temperature of 635\,K~\cite{Kato2002} well above room temperature,
and a predicted high spin polarization of 86\%~\cite{Vaitheeswaran2005}. We
recently have shown that ferromagnetic Sr$_2$CrReO$_6$ grown epitaxially as a
thin film on BaTiO$_3$ substrate exhibits qualitative changes in its magnetic
anisotropy at the BaTiO$_3$ phase transition temperatures~\cite{Czeschka2009}.
Abrupt changes in the coercive field of up to 1.2\,T along with resistance
changes of up to 6.5\% have been observed~\cite{Czeschka2009}. Here, we will
discuss the structural properties of the samples.

\paragraph{Thin Film Deposition.}

Epitaxial Sr$_2$CrReO$_6$ thin films were deposited by laser-MBE with
\textit{in-situ} RHEED \cite{Gross2000,Klein1999} on (001)-oriented BaTiO$_3$
substrates. The thin films with a thickness of approx.~80\,nm were grown in
O$_2$ atmosphere at a pressure of $6.6\times10^{-4}$\,mbar with a laser fluence
of 2\,J/cm$^2$ at the target and a substrate temperature of $700^\circ$C (i.e.
in the cubic, paraelectric phase of BaTiO$_3$). These parameters were found to
be optimal for the growth of Sr$_2$CrReO$_6$ films on SrTiO$_3$
substrates~\cite{Gepraegs2009} and also are appropriate for the growth on
BaTiO$_3$ substrates. After film deposition, the sample is slowly cooled down
to room temperature with the substrate undergoing a phase transition to the
tetragonal and ferroelectric phase. As there is no preferential direction for
the electric polarization, however, the ferroelectric state will decompose into
many different ferroelectric domains as evidenced by X-ray
diffraction~\cite{Czeschka2009}.

\paragraph{Structural Characterization.}

\begin{figure}
    \centering
    \includegraphics[width=\linewidth]{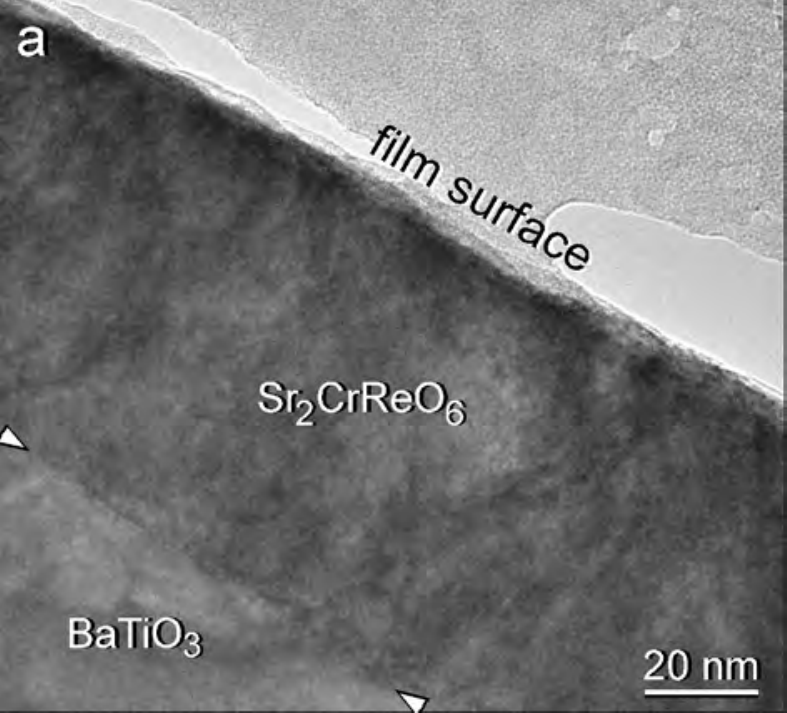}
    \includegraphics[width=\linewidth]{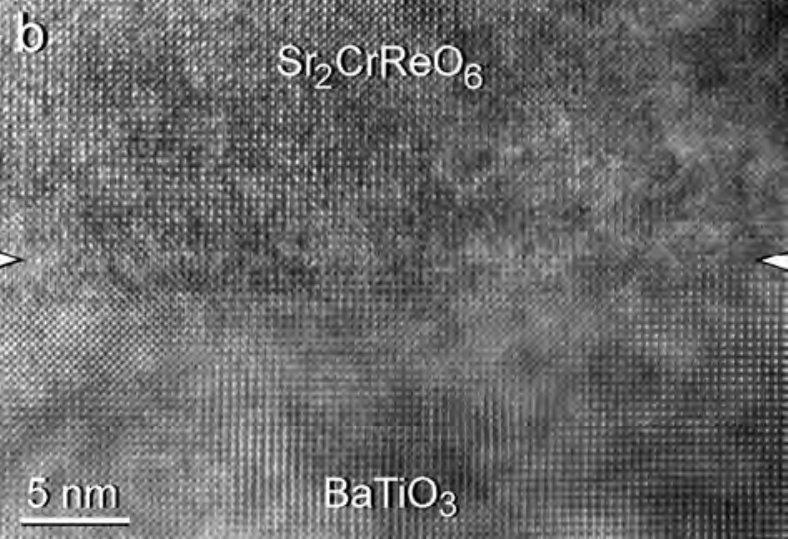}
    \caption{
(a) TEM micrograph of a 81\,nm thin film of Sr$_2$CrReO$_6$ grown on a BaTiO$_3$
substrate. (b) HR-TEM lattice image shows intimately bonded crystals at the
interface. Image contrast in the two perovskite-type oxides proves continuation
of the $A$ and $B$ site metal ion positions and hence direct transition of the
oxygen sub-lattices over wide areas.}
    \label{fig:MF-Sr2CrReO6-TEM}
\end{figure}

High-resolution TEM studies were performed at thin cross-sections of the
Sr$_2$CrReO$_6$ films on BaTiO$_3$ prepared by standard specimen preparation
techniques. Fig.~\ref{fig:MF-Sr2CrReO6-TEM}(a) shows a thin film with
ca.~81\,nm thickness. Electron diffraction reveals (i) a perfect epitaxial
orientation relationship between the double perovskite film and BaTiO$_3$ and
(ii) a large single domain of the tetragonal substrate. However, the difference
between the $a$ and $c$ lattice parameter of tetragonal BaTiO$_3$ as well as the cell
parameter differences in the strained Sr$_2$CrReO$_6$ film (see
Fig.~\ref{fig:MF-Sr2CrReO6-TEM}) are ca.~1\% and are thus too small to be
measured by electron diffraction. The HR-TEM micrograph
(Fig.~\ref{fig:MF-Sr2CrReO6-TEM}(b)) is of the interface region which appears
without any voids or other defects. The lattice mismatch between $c/2$ of
Sr$_2$CrReO$_6$ and $c$ of BaTiO$_3$ can be estimated to ca.~2\%
(Fig.~\ref{fig:MF-Sr2CrReO6-TEM}), however misfit dislocations were not
observed regularly. The lattice image contrast in the Sr$_2$CrReO$_6$ film is
clearly dominated by the ${\rm Re} + {\rm Cr} + 2{\rm O}$ columns ($Z_{\rm Re}
+ Z_{\rm Cr} + 2Z_{\rm O} = 75 + 24 + 16 = 115$) vs.~the Sr columns ($2Z_{\rm
Sr} = 76$) while in BaTiO$_3$ the barium columns ($Z_{\rm Ba} = 56$)
predominantly cause the image contrast. In both crystals, a square array of
dots is produced with spacing of $a$ or $c$ of BaTiO$_3$ and the spacing of
$a/\sqrt{2}$ or $c/2$ of Sr$_2$CrReO$_6$, respectively. Close inspection of the
image contrast reveals a gradual shift of the square dot array along the
interface from BaTiO$_3$ to Sr$_2$CrReO$_6$ by half of the square length
showing the continuation of the $A$ site and $B$ site positions and hence
proving an undisturbed transition of the oxygen sub-lattices of the two
perovskite-type crystals across the interface.

\begin{figure}
 \centering
    \includegraphics[width=\linewidth]{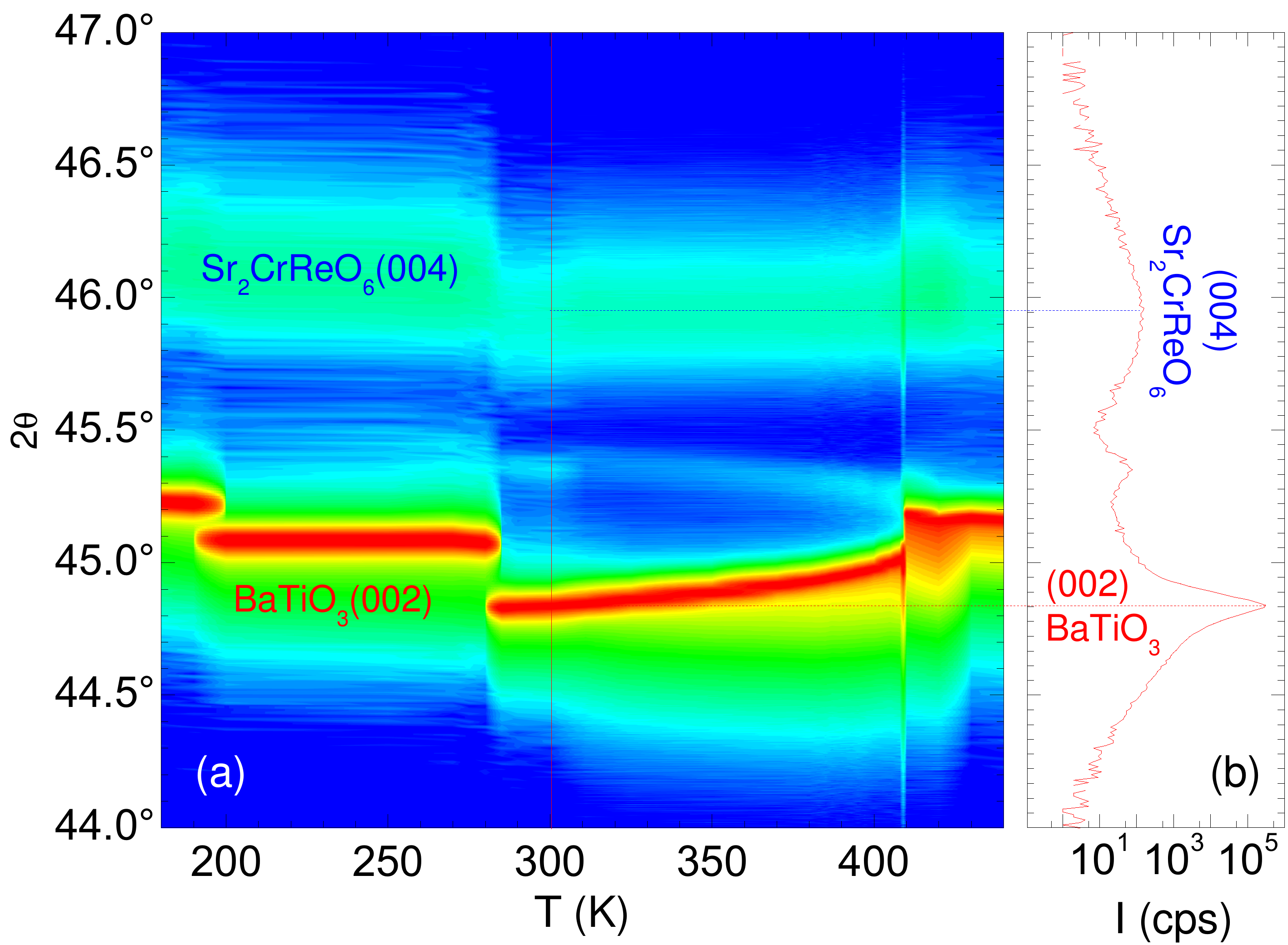}
    \caption{
X-ray diffraction from a 81\,nm thin Sr$_2$CrReO$_6$ film, grown on a BaTiO$_3$
substrate, i.e. the same sample as in Fig~\ref{fig:MF-Sr2CrReO6-TEM}. (a)
Contour plot showing the color coded intensity (red = high, blue = low) from
$\omega$--$2\theta$--scans as a function of temperature. (b) Line scan for $T
= 300$\,K (red vertical line). The reflections from the BaTiO$_3$ substrate and
the Sr$_2$CrReO$_6$ thin film are labeled in red and blue, respectively.}
    \label{fig:MF-Sr2CrReO6-XRD}
\end{figure}

High-resolution X-ray diffraction was performed using Cu K$\alpha_1$ radiation
and a high-resolution four-circle diffractometer (Bruker AXS discover).
Fig.~\ref{fig:MF-Sr2CrReO6-XRD} shows $\omega-2\theta$-($00\ell$)-scans from
the same sample as in Fig.~\ref{fig:MF-Sr2CrReO6-TEM} for different
temperatures between 180\,K and 450\,K. The (004) reflection from the
Sr$_2$CrReO$_6$ thin film and the (002) reflection from the BaTiO$_3$ substrate
can be clearly distinguished. It is obvious that the $2\theta$ angles of both
reflections change abruptly when the BaTiO$_3$ undergoes structural phase
transitions. The line scan at 300\,K (Fig.~\ref{fig:MF-Sr2CrReO6-XRD}(b))
reveals no crystalline parasitic phases in the thin film. A more detailed
picture is obtained from reciprocal space maps showing that the reflection from
the BaTiO$_3$ substrate is split due to the presence of different ferroelectric
domains in the tetragonal phase ($a$ domains: BaTiO$_3$(200/020) and $c$
domains: BaTiO$_3$(002))~\cite{Gepraegs2009}.
From the intensity ratio of the Sr$_2$CrReO$_6$(101) and (404)
reflections (not shown here) the amount of Cr/Re disorder is estimated
to be less than 30\%, i.e. less than 30\% of the Cr$^{3+}$ ions are on Re$^{5+}$ sites
and vice versa \cite{Czeschka2009}.

\begin{figure}
 \centering
    \includegraphics[width=\linewidth]{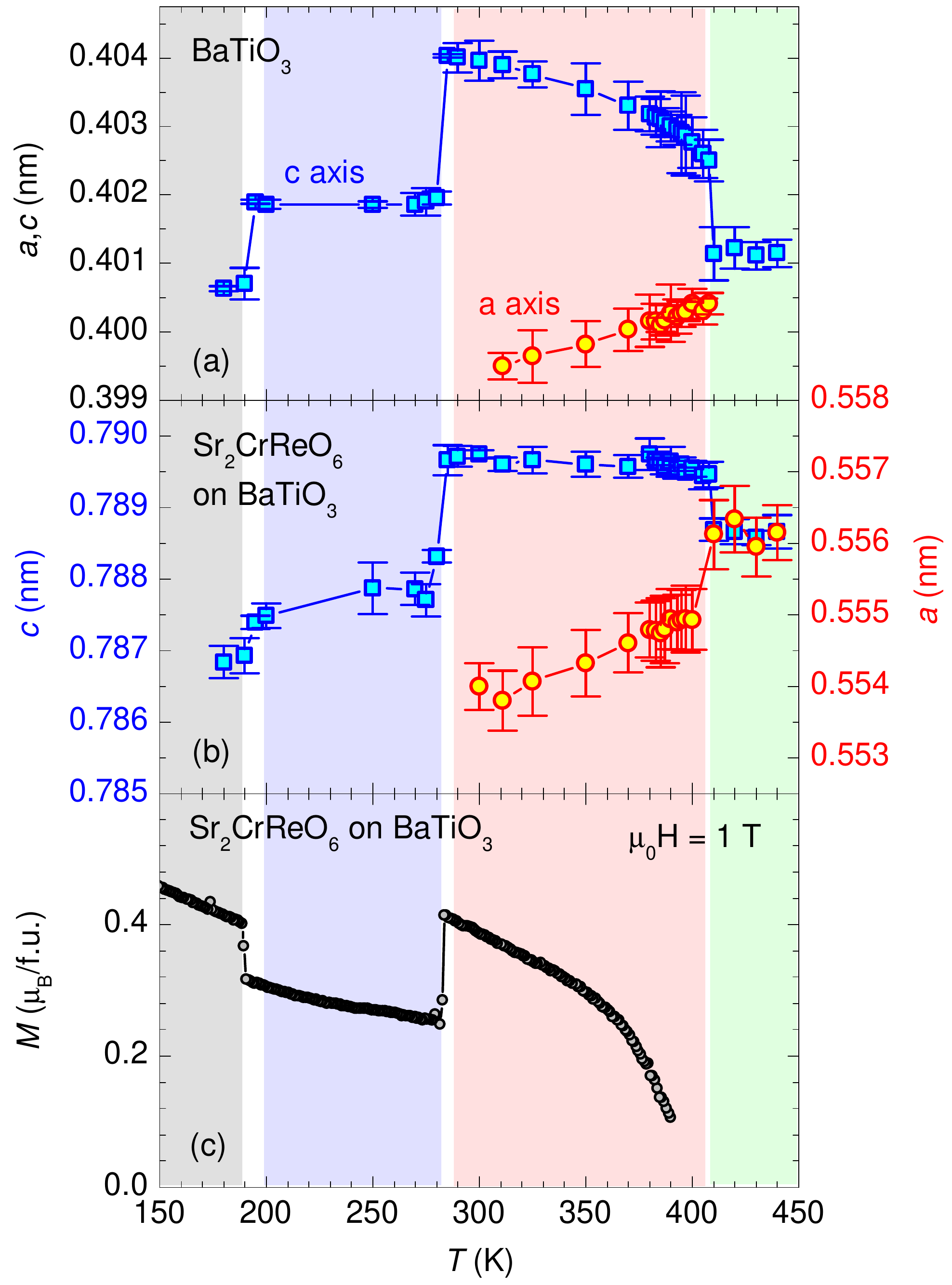}
    \caption{
(a) Lattice parameter $a$ (red circles) and $c$ (blue squares) of the BaTiO$_3$
substrate as a function of temperature, determined by X-ray diffraction. (b)
$a$ (red circles, right scale) and $c$ (blue squares, left scale) for a 81\,nm
thin film of Sr$_2$CrReO$_6$, grown on BaTiO$_3$ substrate, i.e. the same
sample as in Fig.~\ref{fig:MF-Sr2CrReO6-TEM}. (c) Magnetization of a 50\,nm
thin film of Sr$_2$CrReO$_6$, grown on BaTiO$_3$ substrate. The data in (a-c)
were taken while cooling.
The color shaded areas mark the temperature regimes of the crystallographic
phases of BaTiO$_3$ and correspond to the colors used in
Fig.~\ref{fig:MF-BaTiO3}. In (a,b), an electric voltage of 200\,V was applied
to the BaTiO$_3$ substrate, in (c) a magnetic field of 1\,T was applied in the
film plane.}
    \label{fig:MF-Sr2CrReO6-BaTiO3}
\end{figure}

From the observed $2\theta$ angles, we calculate the $a$ and $c$ lattice
parameters for both the substrate and the thin film. The results are displayed
in Fig.~\ref{fig:MF-Sr2CrReO6-BaTiO3}(a,b). All data were obtained with
decreasing temperature. The $c$ axis lattice parameter of our BaTiO$_3$
substrate (Fig.~\ref{fig:MF-Sr2CrReO6-BaTiO3}(a)) nicely follows the behavior
reported for bulk material (see Fig.\ref{fig:MF-BaTiO3}) \cite{Shebanov1981}. It is slightly above
$c=0.401$\,nm in the cubic phase, then increases to $c=0.4025 \ldots 0.404$\,nm
(tetragonal), drops to $c=0.402$\,nm (orthorhombic), and finally becomes $c
\simeq 0.4005$\,nm (rhombohedral). Also the $a$ parameter behaves as expected
in the tetragonal phase. For the Sr$_2$CrReO$_6$ thin film
(Fig.~\ref{fig:MF-Sr2CrReO6-BaTiO3}(b)), the $a$ axis parameter shows the same
temperature dependence as for the substrate. This confirms coherent
growth and biaxial strain transfer from BaTiO$_3$ into the epitaxial film. Also
the $c$ lattice parameter follows the trend of the substrate. In the tetragonal
phase, however, it does not display any temperature dependence, in contrast to
the behavior of BaTiO$_3$. In summary, the structural analysis shows that
the lattice parameters of the Sr$_2$CrReO$_6$ thin film change abruptly at the
BaTiO$_3$ phase transition temperatures.

\paragraph{Magnetic Characterization.}

The magnetic properties were determined by a SQUID magnetometer with a magnetic
field of 1\,T applied in the film plane
(Fig.~\ref{fig:MF-Sr2CrReO6-BaTiO3}(c)). At the BaTiO$_3$ phase transitions, we
observe abrupt changes of the magnetization $M$, which drops at the
tetragonal-orthorhombic transition and increases again when entering the
rhombohedral phase. We note that we observe a small thermal hysteresis due to
the first order structural phase transition of the substrate. A careful
investigation~\cite{Czeschka2009} shows that the values for saturation
magnetization, remanent magnetization, and coercivity clearly differ from each
other for the different structural phases of the substrate. Therefore, we
attribute the observed discontinuities to a change of the magnetic anisotropy
of the thin Sr$_2$CrReO$_6$ layer. They are induced by abrupt changes of the
epitaxial coherency strain due to the structural phase transitions of the
BaTiO$_3$ substrate. We note that the longitudinal resistance also shows abrupt
changes at the BaTiO$_3$ phase transition temperatures as evidenced by
electrical transport measurements in photolithographically patterned Hall
bars~\cite{Czeschka2009}. These ``jumps'' are as large as 5.0\%
(rhombohedral/orthorhombic) and 4.2\% (orthorhombic/tetragonal) and therefore
by a factor of 5 larger than what is expected from a pure geometric effect due
to a change in the spatial dimensions of the Hall bar~\cite{Czeschka2009}.

\paragraph{Summary.}

We have grown heteroepitaxial hybrid structures of Sr$_2$CrReO$_6$/BaTiO$_3$ by
laser-MBE. Due to coherent growth, the in-plane lattice constants of the
Sr$_2$CrReO$_6$ film follow those of the BaTiO$_3$ substrate as a function of
temperature. In particular, the epitaxial coherency strain in the
Sr$_2$CrReO$_6$ film abruptly changes at the structural phase transitions of
BaTiO$_3$. The magnetization of the ferromagnetic Sr$_2$CrReO$_6$ film together
with its resistivity~\cite{Gepraegs2009} shows large step-like variations at
the corresponding phase transition temperatures. This demonstrates a strong
coupling of the magnetic properties to elastic distortions, i.e. a large
magnetoelastic coupling. Taken together, a strong coupling between the
ferroelectric order parameter in BaTiO$_3$ and the ferromagnetic order
parameter in Sr$_2$CrReO$_6$ has been shown. That is, the
Sr$_2$CrReO$_6$/BaTiO$_3$ heterostructure can be considered as a
magnetoelectric multiferroic.

Our observations suggest that the system Sr$_2$CrReO$_6$/ BaTiO$_3$ in
particular, and other suitable ferromagnetic/fer\-ro\-electric heterostructures in
general, are promising candidates for the manipulation of magnetic material
properties by an electric field at room temperature. In such structures the
electric field is applied to the ferroelectric material, thereby causing
changes of the lattice constants by the inverse piezoelectric effect or by
switching the ferroelectric domain structure (e.g. from $c$-domains to
$a$-domains and vice versa in BaTiO$_3$). The elastic deformation is
transferred into elastic strain in the ferromagnetic film, causing a controlled
variation of the magnetic properties via magnetoelastic coupling. Altogether,
the coupling between an electric field and the magnetic sample properties, i.e. a
magnetoelectric coupling, is realized.

\section{Conclusion.}
 \label{sec:Conclusion}

The combination of several physical properties such as superconductivity,
ferro- and antiferromagnetism, ferroelectricity, or even multiferroicity in
artificial heterostructures paves the way to a rich variety of interesting new
physics and novel concepts in condensed matter physics. Regarding materials,
oxide thin films and artificial heterostructures are at the forefront of this
rapidly emerging field of materials science. On the one hand, they have
outstanding physical properties and, on the other hand, there is enormous
progress in oxide thin film technology. We have shown that the combination of
ferromagnetic, semiconducting, metallic, and dielectric materials properties in
thin films and artificial heterostructures is promising. Using laser molecular
beam epitaxy, we have fabricated heterostructures for the realization of
oxide-based ferromagnetic tunnel junctions, transition metal-doped semiconductors,
intrinsic multiferroics, and artificial ferroelectric/ferromagnetic
heterostructures. The latter have been used to study magnetoelastic coupling,
forming the basis of the promising field of spin-mechanics.

Artificial oxide heterostructures are particularly promising for the
realization of materials with improved and new functionalities and novel device
concepts. Hereby, interface and surface effects play a crucial role. Due to the
complexity of the involved oxide materials, the rich variety of physics
resulting from band bending effects, magnetic exchange, or elastic coupling at
interfaces in heterostructures is far from being understood and needs further
detailed studies. Moreover, intermixing of the different atomic species
deposited in multilayer structures plays an important role and might influence
or even dominate the overall physical properties. Therefore, a careful
structural and element-specific investigation of the artificial material system
or device structure on an atomic scale is highly necessary.

\begin{acknowledgement}
Financial support by the Deutsche Forschungsgemeinschaft via the priority
program 1157 (Project Nos. GR 1132/13 \& MA 1020/12), Project No. GO 944/3,
and the Excellence Cluster
\textit{Nanosystems Initiative Munich (NIM)}, as well as by the ESRF (Project No.
HE-2089) is gratefully acknowledged. We thank Andreas Erb for the preparation
of the polycrystalline target materials for the pulsed laser deposition process,
Thomas Brenninger for continuous technical support, and Deepak Venkateshvaran
for valuable discussions.

\end{acknowledgement}

%

\begin{thebibliography}{[1]}

\bibitem{Blamire2009}
    M. G. Blamire, J. L. MacManus-Driscoll, N. D. Mathur, and Z. H. Barber, Adv. Mater. \textbf{21}, 3827 (2009).
\bibitem{Johnsson2008}
    M. Johnsson and P. Lemmens, J. Phys.: Condens. Matter \textbf{20}, 264001 (2008).
\bibitem{Bednorz1986}
    J. G. Bednorz and K. A. M\"{u}ller, Z. Phys. B \textbf{64}, 189 (1986).
\bibitem{Jia1986}
    C.-L. Jia, V. Nagarajan, J.-Q. He, L. Houben, T. Zhao, R. Ramesh, K. Urban, R. Waser, Nature Mater. \textbf{6}, 64 (2007).
\bibitem{Zimmermann2001}
    F. Zimmermann, M. Voigts, C. Weil, R. Jakoby, P. Wang, W. Menesklou, E. Ivers-Tiff\'{e}e, J. Europ. Cer. Soc. \textbf{21}, 2019 (2001).
\bibitem{Helmolt1993}
    R. von Helmholt, J. Wecker, B. Holzapfel, L. Schultz, and K. Samwer, Phys. Rev. Lett. \textbf{71}, 2331 (1993).
\bibitem{Jin1994}
    S. Jin, T. H. Tiefel, M. McCormack, R. A. Fastnacht, R. Ramesh, and L. H. Chen, Science \textbf{264}, 413 (1994).
\bibitem{Spaldin2005}
    N.A. Spaldin and M. Fiebig, Science \textbf{309}, 391 (2005).
\bibitem{Fiebig2005}
    M. Fiebig, J. Phys. D \textbf{38}, R123-R152 (2005).
\bibitem{Ohno1996}
    H. Ohno, A. Shen, F. Matsukura, A. Oiwa, A. Endo, S. Katsumoto, and Y. Iye, Appl. Phys. Lett. \textbf{69}, 363 (1996).
\bibitem{Dietl2000}
    T. Dietl, H. Ohno, F. Matsukara, J. Cibert, and D. Ferrand, Science \textbf{287}, 1019 (2000).
\bibitem{Venkatesan2004}
    M. Venkatesan, C.B. Fitzgerald, J.-G. Lunney, and J.M.D. Coey, Phys. Rev. Lett. \textbf{93}, 177206 (2004).
\bibitem{Gross2000} R. Gross, J. Klein, B. Wiedenhorst, C. H\"{o}fener, U. Schoop,
    J. B. Philipp, M. Schonecke, F. Herbstritt, L. Alff, Yafeng Lu, A. Marx, S.
    Schymon, S. Thienhaus, and W. Mader, Proc. SPIE \textbf{4058}, 278 (2000).
\bibitem{Gupta1990} A. Gupta, R. Gross, E. Olsson, A. Segm\"{u}ller, G. Koren, C.
    C. Tsuei, Phys. Rev. Lett. \textbf{64}, 3191 (1990).
\bibitem{Klein1999}
    J. Klein, C. H\"{o}fener, L. Alff, and R. Gross, Supercond. Sci. Technol. \textbf{12}, 1023 (1999).
\bibitem{Klein2000} J. Klein, C. H\"{o}fener, L. Alff, R. Gross, J. Magn. Magn.
    Mater. \textbf{211}, 9 (2000).
\bibitem{Reisinger2003a} D.~Reisinger, B. Blass, J. Klein, J.~B.~Philipp,
    M.~Schonecke, A.~Erb, L.~Alff, and R.~Gross, Appl. Phys. A \textbf{77}, 619--621 (2003).
\bibitem{Ohtomo2004}
    A. Ohtomo, H. Y. Hwang, Nature \textbf{427}, 423 (2004).
\bibitem{Reyren2007}
    N. Reyren, S. Thiel, A. D. Caviglia, L. Fitting Kourkoutis, G. Hammerl, C. Richter, C. W. Schneider, T. Kopp, A.-S. R\"{u}etschi, D. Jaccard, M. Gabay, D. A. Muller, J.-M. Triscone, and J. Mannhart, Science \textbf{317}, 1196 (2007).
\bibitem{Zhuravlev2005} M.Ye. Zhuravlev, R.F. Sabirianov, S.S. Jaswal, and
    E.Y. Tsymbal, Phys. Rev. Lett. \textbf{940}, 246802 (2005).
\bibitem{Tsymbal2006} E.Y. Tsymbal, H. Kohlstedt, Science \textbf{313},
    181--83 (2006).
\bibitem{Gross1990} R. Gross, A. Gupta, E. Olsson, A. Segm\"{u}ller, G. Koren,
    Appl. Phys. Lett. \textbf{57}, 203 (1990).
\bibitem{Wiedenhorst1999} B.~Wiedenhorst, C.~H\"{o}fener, Yafeng~Lu, J.~Klein,
    L.~Alff, R.~Gross, B. H.~Freitag, W.~Mader, Appl. Phys. Lett. \textbf{74}, 3636 (1999).
\bibitem{Wiedenhorst2000} B. Wiedenhorst, C. H\"{o}fener, Yafeng Lu, J. Klein,
    M. S. R. Rao, H. Freitag, W. Mader, L. Alff, R. Gross, J. Magn. and Magn. Mat. \textbf{211}, 16
    (2000).
\bibitem{Lu2000} Yafeng Lu, J. Klein, C. H\"{o}fener, B. Wiedenhorst, J. B.
    Philipp, F. Herbstritt, L. Alff, and R. Gross, Phys. Rev. B \textbf{62}, 15806 (2000).
\bibitem{Lu2005} Yafeng Lu, J. Klein, F. Herbstritt, J.B. Philipp, A. Marx,
    L. Alff, R. Gross, phys. stat. sol. (b) \textbf{242}, 1545--1560 (2005).
\bibitem{Goennenwein2008} S.T.B. Goennenwein, M. Althammer, C. Bihler, A.
    Brandlmaier, S. Gepr\"{a}gs, M. Opel, R. Gross, W. Schoch, W. Limmer, H. Huebl,
    M. S. Brandt, phys. stat. sol. (RRL) \textbf{2}, 96--98 (2008).
\bibitem{Brandlmaier2008}
    A. Brandlmaier, S. Gepr\"{a}gs, M. Weiler, A. Boger, M. Opel, H. Huebl, C. Bihler, M.S. Brandt, B. Botters, D. Grundler, R. Gross, S.T.B. Goennenwein, Phys. Rev. B \textbf{77}, 104445 (2008).
\bibitem{Weiler2009}
    M. Weiler, A. Brandlmaier, S. Gepr\"{a}gs, M. Althammer, M. Opel, C. Bihler, H. Huebl, M. S. Brandt, R. Gross, and S. T. B. Goennenwein, New J. Phys. \textbf{11}, 013021 (2009).
\bibitem{Choi2004} K. J. Choi, M. Biegalski, Y. L. Li,  A. Sharan, J.
    Schubert, R. Uecker, P. Reiche, Y. B. Chen, X. Q. Pan, V. Gopalan, L.-Q. Chen, D.
    G. Schlom, C. B. Eom, Science \textbf{306}, 1005 (2004).
\bibitem{Bousquet2008} E. Bousquet, M. Dawber, N. Stucki, C. Lichtensteiger,
    P. Hermet, S. Gariglio, J.-M. Triscone, P. Ghosez, Nature \textbf{452}, 732
    (2008).
\bibitem{Sai2000}
    N. Sai, B. Meyer, and D. Vanderbilt, Phys. Rev. Lett. \textbf{84}, 5636 (2000).
\bibitem{Lee2005}
    H. N. Lee, H. M. Christen, M. F. Chisholm, C. M. Rouleau, and D. H. Loundes, Nature \textbf{433}, 395 (2005).
\bibitem{Warusawithana2003}
    M. P. Warusawithana, E. V. Colla, J. N. Eckstein, and M. B. Weissman, Phys. Rev. Lett. \textbf{90}, 036802 (2003).
\bibitem{Kanki2006}
    T. Kanki, H. Tanaka, and T. Kawai, Appl. Phys. Lett. \textbf{89}, 242506 (2006).
\bibitem{Salvador1999}
    P. A. Salvador, A.-M. Haghiri-Gosnet, B. Mercey, M. Hervieu, and B. Raveau, Appl. Phys. Lett. \textbf{75}, 2638 (1999).
\bibitem{Neaton2003}
    J. B. Neaton and K. M. Rabe, Appl. Phys. Lett. \textbf{82}, 1586 (2003).
\bibitem{Pertsev2000}
    N. A. Pertsev, A. K. Tagantsev and N. Setter, Phys. Rev. B \textbf{61}, R825 (2000).
\bibitem{Okamoto2004}
    S. Okamoto and A. J. Millis, Nature \textbf{428}, 630 (2004).
\bibitem{Junquera2003}
    J. Junquera and P. Ghosez, Nature \textbf{422}, 506 (2003).
\bibitem{Duan2006}
    C.-G. Duan, S. S. Jaswal and E. Y. Tsymbal, Phys. Rev. Lett. \textbf{97}, 047201 (2006).
\bibitem{Stengel2006}
    M. Stengel and N. A. Spaldin, Nature \textbf{443}, 679 (2006).


\bibitem{Prinz1998}
    G. A. Prinz, Science \textbf{282}, 1660 (1998).
\bibitem{Wolf2001} S.A. Wolf, D. D. Awschalom, R. A. Buhrman, J. M. Daughton,
    S. von Molnar, M. L. Roukes, A. Y. Chtchelkanova, and D. M. Treger, 
    Science {\bf 294}, 1488 - 1495 (2001).
\bibitem{Zutic2004}
    Igor \v{Z}uti\'{c}, J. Fabian, S. Das Sarma, Rev. Mod. Phys. \textbf{76}, 323 (2004).
\bibitem{Coey2003} J.M.D. Coey and C.L. Chien, MRS Bull. \textbf{28}, 720
    (2003).
\bibitem{Gross2006}
    R. Gross, in: Nanoscale Devices -- Fundamentals and Applications, edited by R. Gross, A. Sidorenko, and L. Tagirov (Springer, Berlin, 2006), pp. 49-110.
\bibitem{Philipp2001}
    J. B. Philipp, D. Reisinger, M. Schonecke, A. Marx, A. Erb, L. Alff, and R. Gross, Appl. Phys. Lett. \textbf{79}, 3654 (2001).
\bibitem{Philipp2003a}
    J. B. Philipp, D. Reisinger, M. Schonecke, M. Opel, A. Marx, A. Erb, L. Alff, and R. Gross, J. Appl. Phys. \textbf{93}, 6853 (2003).
\bibitem{Majewski2005b}
    P. Majewski, S. Gepr\"{a}gs, A. Boger, M. Opel, L. Alff, and R. Gross, J. Magn. Magn. Mater. \textbf{290-291}, 1154 (2005).
\bibitem{Nielsen2008}
    A. Nielsen, A. Brandlmaier, M. Althammer, W. Kaiser, M. Opel, J. Simon, W. Mader, S. T. B. Goennenwein, and R. Gross, Appl. Phys. Lett. \textbf{93}, 162510 (2008).
\bibitem{Majewski2005a}
    P. Majewski, S. Gepr\"{a}gs, A. Boger, M. Opel, A. Erb, L. Alff, R. Gross, G. S. Vaitheeswaran, V. Kanchana, A. Delin, F. Wilhelm, and A. Rogalev, Phys. Rev. B \textbf{72}, 132402 (2005).
\bibitem{Majewski2005c}
    P. Majewski, S. Gepr\"{a}gs, O. Sanganas, M. Opel, R. Gross, F. Wilhelm, A. Rogalev, and L. Alff, Appl. Phys. Lett. \textbf{87}, 202503 (2005).
\bibitem{Gepraegs2006}
    S. Gepr\"{a}gs, P. Majewski, C. Ritter, L. Alff, and R. Gross, J. Appl. Phys. \textbf{99}, 08J102 (2006).
\bibitem{Venkateshvaran2009}
    D. Venkateshvaran, M. Althammer, A. Nielsen, S. Gepr\"{a}gs, M. S. R. Rao, S. T. B. Goennenwein, M. Opel, and R. Gross, Phys. Rev. B \textbf{79}, 134405 (2009).
\bibitem{Philipp2003b}
    J. B. Philipp, J. Klein, S. Afilal, C. Recher, T. Walther, W. Mader, M. Schmid, R. Suryanarayanan, L. Alff, and R. Gross, Phys. Rev. B \textbf{68}, 144431 (2003).
\bibitem{Philipp2002} J. B. Philipp, L. Alff, A. Marx, R. Gross, Phys. Rev. B
    \textbf{66}, 224417 (2002).
\bibitem{Philipp2004} J. B. Philipp, P. Majewski, D. Reisinger, S. Gepr\"{a}gs, M.
    Opel, A. Erb, L. Alff, R. Gross, Acta Phys. Pol. A \textbf{105}, 7--26 (2004).
\bibitem{Opel2008}
    M. Opel, K.-W. Nielsen, S. Bauer, S.T.B. Goennenwein, J.C.Cezar, D. Schmeisser, J. Simon, W. Mader, and R. Gross, Eur. Phys. J. B \textbf{63}, 437 (2008).
\bibitem{Gepraegs2007}
    S. Gepr\"{a}gs, M. Opel, S.T.B. G\"{o}nnenwein, and R. Gross, Phil. Mag. Lett. \textbf{87}, 141 (2007).


\bibitem{Akerman2005}
    J. {\AA}kerman, Science \textbf{308}, 508 (2005).
\bibitem{Ney2003}
    A. Ney, C. Pampuch, R. Koch, and K. H. Ploog, Nature \textbf{425}, 485 (2003).
\bibitem{Julliere1975}
    M. Julli\`{e}re, Phys. Lett. A \textbf{54}, 225 (1975).
\bibitem{Moodera1999}
    J. S. Moodera, J. Nassar, and G. Mathon, Annu. Rev. Mater. Sci. \textbf{29}, 381 (1999).
\bibitem{Butler2001}
    W. H. Butler, X.-G. Zhang, T. C. Schulthess, and J. M. MacLaren, Phys. Rev. B \textbf{63}, 054416 (2001).
\bibitem{Mathon2001}
    J. Mathon and A. Umerski, Phys. Rev. B \textbf{63}, 220403 (2001).
\bibitem{Yuasa2004}
    S. Yuasa, T. Nagahama, A. Fukushima, Y. Suzuki, and K. J. Ando,  Nature Mater. \textbf{3}, 868 (2004).
\bibitem{Parkin2004}
    S. S. P. Parkin, C. Kaiser, A. Panchula, P. M. Rice, B. Hughes, M. Samant, and S.-H. Yang, Nature Mater. \textbf{3}, 862 (2004).
\bibitem{Ikeda2008}
    S. Ikeda, J. Hayakawa, Y. Ashizawa, Y. M. Lee, K. Miura, H. Hasegawa, M. Tsunoda, F. Matsukura, and H. Ohno, Appl. Phys. Lett. \textbf{93}, 082508 (2008).
\bibitem{Zhang1991}
    Z. Zhang and S. Satpathy, Phys. Rev. B, \textbf{44}, 13319 (1991).
\bibitem{Gorter1955}
    E. W. Gorter, Proc. IRE \textbf{43}, 1945 (1955).
\bibitem{Dedkov2002}
    Yu. S. Dedkov, U. R\"{u}diger, and G. G\"{u}ntherodt, Phys. Rev. B \textbf{65}, 064417 (2002).
\bibitem{Fonin2008} M. Fonin, Yu. S. Dedkov, R. Pentcheva, U. R\"{u}diger, and
    G. G\"{u}ntherodt, J. Phys.: Condens. Matter \textbf{20}, 142201 (2008).
\bibitem{Aoshima2003}  K. Aoshima and S. X. Wang, J. Appl. Phys.
    \textbf{93}, 7954 (2003).
\bibitem{vanderZaag2000} P. J. van der Zaag, P. J. H. Bloemen, J. M.
    Gaines, R. M. Wolf, P. A. A. van der Heijden, R. J. M. van de Veerdonk, and
    W. J. M. de Jonge; J. Magn. Magn. Mater. \textbf{211}, 301 (2000).
\bibitem{Li1998} X. W. Li, A. Gupta, G. Xiao, W. Qian, and V. P. Dravid,
    Appl. Phys. Lett. \textbf{73}, 3282 (1998).
\bibitem{Park2003} C. Park, Y. Shi, Y. Peng, K. Barmak, J.-G. Zhu, D. E.
    Laughlin, and R. M. White, IEEE Transactions on Magnetics \textbf{39}, 2806
    (2003).
\bibitem{Matsuda2002} H. Matsuda, M. Takeuchi, H. Adachi, M. Hiramoto, N.
    Matsukawa, A. Odagawa, K. Setsune, and H. Sakakima, Jpn. J. Appl. Phys.
    \textbf{41}, L387 (2002).
\bibitem{Seneor1999} P. Seneor, A. Fert, J.-L. Maurice, F. Montaigne, F.
    Petroff, and A. Vaures, Appl. Phys. Lett. \textbf{74}, 4017 (1999).
\bibitem{Reisinger2003b}
    D. Reisinger, M. Schonecke, T. Brenninger, M. Opel, A. Erb, L. Alff, and R. Gross, J. Appl. Phys. \textbf{94}, 1857 (2003).
\bibitem{Alff1992} L. Alff, G. Fischer, R. Gross, F. Kober, K. D. Husemann,
    A. Beck, T. Nissel, C. Burckhardt, F. Schmidl, Physica C \textbf{200}, 277 (1992).
\bibitem{Reisinger2004}
    D. Reisinger, P. Majewski, M. Opel, L. Alff, and R. Gross, Appl. Phys. Lett. \textbf{85}, 4980 (2004).
\bibitem{Venkateshvaran2008}
    D. Venkateshvaran, W. Kaiser, A. Boger, M. Althammer, M. S. R. Rao, S. T. B. Goennenwein, M. Opel, and R. Gross, Phys. Rev. B \textbf{78}, 092405 (2008).
\bibitem{Verwey1939}
    E.J.W. Verwey, Nature \textbf{144}, 327 (1939).
\bibitem{Shepherd1985}
    J.P. Shepherd, R. Aragon, J.W. Koenitzer, and J.M. Honig, Phys. Rev. B \textbf{32}, 1818 (1985).
\bibitem{Chen2000}
    E.Y. Chen, R. Whig, J.M. Slaughter, D. Cronk, J. Goggin, G. Steiner, and S. Tehrani, J. Appl. Phys. \textbf{87}, 6061 (2000).
\bibitem{Gao2009}
    L. Gao, X. Jiang, P.M. Rice, and S.S.P. Parkin, Appl. Phys. Lett. \textbf{95}, 122503 (2009).
\bibitem{deTeresa1999}
    J.M. De Teresa, A. Barth\'{e}l\'{e}my, A. Fert, J.P. Contour, R. Lyonnet, F. Montaigne, P. Seneor, and A. Vaur\`{e}s, Phys. Rev. Lett. \textbf{82}, 4288 (1999).
\bibitem{Hu2002}
    G. Hu and Y. Suzuki, Phys. Rev. Lett. \textbf{89}, 276601 (2002); G. Hu, R. Chopdekar, and Y. Suzuki, J. Appl. Phys. \textbf{93}, 7516 (2003).
\bibitem{Hoefener2000} C.~H\"{o}fener, J.B.~Philipp, J.~Klein, L.~Alff, A.~Marx, B.
    B\"{u}chner, and R.~Gross, Europhys. Lett. \textbf{50}, 681 (2000).
\bibitem{Klein1999a} J. Klein, C. H\"{o}fener, S. Uhlenbruck, L. Alff, B. B\"{u}chner,
    R. Gross, Europhys. Lett. \textbf{47}, 371 (1999).
\bibitem{Alvarado1992}
    S.F. Alvarado and P. Renaud, Phys. Rev. Lett. \textbf{68}, 1387 (1992).
\bibitem{Bataille2007}
    A.M. Bataille, R. Mattana, P. Seneor, A. Tagliaferri, S. Gota, K. Bouzehouane, C. Deranlot, M.-J. Guittet, J.-B. Moussy, C. de Nada\"{\i}, N.B. Brookes, F. Petroff, and M. Gautier-Soyer,
    J. Magn. Magn. Mater. \textbf{316}, e963 (2007).
\bibitem{Moodera1996}
    J.S. Moodera, L.R. Kinder, J. Nowak, P. LeClair, and R. Meservey, Appl. Phys. Lett. \textbf{69}, 708 (1996).
\bibitem{Pedersen1967}
    R.J. Pedersen and F.L. Vernon Jr., Appl. Phys. Lett. \textbf{10}, 29 (1967).
\bibitem{Veerdonk1997}
    R.J.M. van de Veerdonk, J. Nowak, R. Meservey, J.S. Moodera, and W.J.M. de Jonge, Appl. Phys. Lett. \textbf{71}, 2839 (1997).
\bibitem{Kimoto1997}
    K. Kimoto, T. Sekiguchi, and T. Aoyama, J. Electron Microsc. \textbf{46}, 369 (1997).
\bibitem{Walther2003}
    T. Walther, Ultramicroscopy \textbf{96}, 401 (2003).

\bibitem{Behan2008}
    A. J. Behan, A. Mokhtari, H. J. Blythe, D. Score, X-H. Xu, J. R. Neal, A. M. Fox, and G. A. Gehring, Phys. Rev. Lett. \textbf{100}, 047206 (2008).
\bibitem{Lee2009}
    H.-J. Lee, E. Helgren, and F. Hellman, Appl. Phys. Lett. \textbf{94}, 212106 (2009).
\bibitem{Chambers2006}
    S. A. Chambers, T. C. Droubay, C. M. Wang, K. M. Rosso, S. M. Heald, D. A. Schwartz, K. R. Kittilstved, and D. R. Gamelin, Materials Today \textbf{9(11)}, 28 (2006).
\bibitem{Ney2010}
    A. Ney, M. Opel, T.C. Kaspar, V. Ney, S. Ye, K. Ollefs, T. Kammermeier, S. Bauer, K.-W. Nielsen, S.T.B. Goennenwein, M.H. Engelhard, S. Zhou, K. Potzger, J. Simon, W. Mader, S.M. Heald, J.C. Cezar, F. Wilhelm, A. Rogalev, R. Gross, and S.A. Chambers, New J. Phys. \textbf{12}, 013020 (2010).
\bibitem{Ahlers2006}
    S. Ahlers, D. Bougeard, N. Sircar, G. Abstreiter, A. Trampert, M. Opel, and R. Gross, Phys. Rev. B \textbf{74}, 214411 (2006).
\bibitem{Jaeger2006}
    C. Jaeger, C. Bihler, T. Vallaitis, S.T.B. Goennenwein, M. Opel, R. Gross, and M.S. Brandt, Phys. Rev. B \textbf{74}, 045330 (2006).
\bibitem{Mamiya2006}
    K. Mamiya, T. Koide, A. Fujimori, H. Tokano, H. Manaka, A. Tanaka, H. Toyosaki, T. Fukumura, and M. Kawasaki, Appl. Phys. Lett. \textbf{89}, 062506 (2006).
\bibitem{Kobayashi2005}
    M. Kobayashi, Y. Ishida, J. Hwang, T. Mizokawa, A. Fujimori, K. Mamiya, J. Okamoto, Y. Takeda, T. Okane, Y. Saitoh,
    Y. Muramatsu, A. Tanaka, H. Saeki, H. Tabata, and T. Kawai, Phys. Rev. B \textbf{72}, 201201 (2005).
\bibitem{Ney2008b}
    A. Ney, T. Kammermeier, V. Ney, S. Ye, K. Ollefs, E. Manuel, S. Dhar, K. H. Ploog, E. Arenholz, F. Wilhelm, and A. Rogalev, Phys. Rev. B \textbf{77}, 233308 (2008).
\bibitem{Schmidt2000}
    G. Schmidt, D. Ferrand, L. W. Molenkamp, A. T. Filip, and B. J. van Wees, Phys. Rev. B \textbf{62}, 4790 (2000).

\bibitem{Eerenstein2006}
    W. Eerenstein, N. D. Mathur, and J. F. Scott, Nature \textbf{442}, 759 (2006).
\bibitem{Ramesh2007}
    R. Ramesh and N. A. Spaldin, Nature Mater. \textbf{6}, 21 (2007).
\bibitem{Hill2000}
    N.A. Hill, J. Phys. Chem. B \textbf{104}, 6694 (2000).
\bibitem{Kimura2003}
    T. Kimura, S. Kawamoto, I. Yamada, M. Azuma, M. Takano, and Y. Tokura, Phys. Rev. B \textbf{67}, 180401 (2003).
\bibitem{Kiselev1963}
    S. V. Kiselev, R. P. Ozerov, and G. S. Zhdanov, Sov. Phys. Dokl. \textbf{7}, 742 (1963).
\bibitem{Smolenskii1961}
    G. A. Smolenskii, V. A. Isupov, A. I. Agranovskaya, and N. N. Krainik, Sov. Phys. Solid State \textbf{2}, 2651 (1961).
\bibitem{Bea2006}
    H. B\'{e}a, M. Bibes, S. Cherifi, F. Nolting, B. Warot-Fonrose, S. Fusil, G. Herranz, C. Deranlot, E. Jacquet, K. Bouzehouane, and A. Barth\'{e}l\'{e}my, Appl. Phys. Lett. \textbf{89}, 242114 (2006).
\bibitem{Bea2008}
    H. B\'{e}a, M. Bibes, F. Ott, B. Dup\'{e}, X.-H. Zhu, S. Petit, S. Fusil, C. Deranlot, K. Bouzehouane, and A. Barth\'{e}l\'{e}my, Phys. Rev. Lett. \textbf{100}, 017204 (2008).
\bibitem{Bibes2008}
    M. Bibes and A. Barth\'{e}l\'{e}my, Nature Mater. \textbf{7}, 425 (2008).
\bibitem{Chu2008}
    Y.-H. Chu, L. W. Martin, M. B. Holcomb, M. Gajek, S.-J. Han, Q. He, N. Balke, C.-H. Yang, D. Lee, W. Hu, Q. Zhan, P.-L. Yang, A. Fraile-Rodr\'{\i}guez, A. Sscholl, S. X. Wang, and R. Ramesh, Nature Mater. \textbf{7}, 478 (2008).
\bibitem{Wang2003}
    J. Wang, J. B. Neaton, H. Zheng, V. Nagarajan, S. B. Ogale, B. Liu, D. Viehland, V. Vaithynathan, D. G. Schlom, U. V. Waghmare, N. A. Spaldin, K. M. Rabe, M. Wuttig, and R. Ramesh, Science \textbf{299}, 1719 (2003).
\bibitem{Eerenstein2005}
    W. Eerenstein, F. D.Morrison, J. Dho,M. G. Blamire, J. F. Scott, N. D.Mathur, Science \textbf{307}, 1203a (2005).
\bibitem{Wang2005}
    J.Wang, A. Scholl, H. Zheng, S. B.Ogale, D. Viehland, D. G. Schlom, N.A. Spaldin, K.M. Rabe, M. Wuttig, L. Mohaddes, J. Neaton, U.Waghmare, T. Zhao, R. Ramesh, Science \textbf{307}, 1203b (2005).
\bibitem{Bea2005}
    H. B\'{e}a, M. Bibes, A. Barth\'{e}l\'{e}my, K. Bouzehouane, E. Jacquet, A. Khodan, J.-P. Contour, S. Fusil, F. Wyczisk, A. Forget, D. Lebeugle, D. Colson, and M. Viret, Appl. Phys. Lett. \textbf{87}, 072508 (2005).
\bibitem{Bea2009}
    H. B\'{e}a, B. Dup\'{e}, S. Fusil, R. Mattana, E. Jacquet, B. Warot-Fonrose, F. Wilhelm, A. Rogalev, S. Petit, V. Cros, A. Anane, F. Petroff, K. Bouzehouane, G. Geneste,  B. Dkhil, S. Lisenkov, I. Ponomareva, L. Bellaiche, M. Bibes, and A. Barth\'{e}l\'{e}my, Phys. Rev. Lett. \textbf{102}, 217603 (2009).
\bibitem{Ederer2005a}
    C. Ederer and N.A. Spaldin, Phys. Rev. B \textbf{71}, 060401 (2005).
\bibitem{Lu2009}
     J. Lu, A. G\"{u}nther, F. Schrettle, F. Mayr, S. Krohns, P. Lunkenheimer, A. Pimenov, V. D. Travkin, A. A. Mukhin, A. Loidl, Eur. Phys. J. B \textbf{75}, 451 (2010); see also: J. Lu, M. Schmidt, P. Lunkenheimer, A. Pimenov, A. A. Mukhin, V. D. Travkin, and A. Loidl, J. Phys.: Conf. Ser. \textbf{200}, 012106 (2010).
\bibitem{Sugawara1968}
    F. Sugawara, S. Iiida, Y. Syono, and S. I. Akimoto, J. Phys. Soc. Jpn. \textbf{25}, 1553 (1968).
\bibitem{Murakami2006}
    M. Murakami, S. Fujino, S. H. Lim, C. J. Long, L. G. Salamanca-Riba, M. Wuttig, I. Takeuchi, V. Nagarajan, and A. Varatharajan, Appl. Phys. Lett. \textbf{88}, 152902 (2006).
\bibitem{Kim2006}
    D. H. Kim, H. N. Lee, M. Varela, and H. M. Christen, Appl. Phys. Lett. \textbf{89}, 162904 (2006).
\bibitem{Niitaka2004}
    S. Niitaka, M. Azuma, M. Takano, E. Nishibori, M. Tanaka, and M. Sakata, Solid State Ionics \textbf{172}, 557 (2004).
\bibitem{Ederer2005b}
    C. Ederer and N.A. Spaldin, Phys. Rev. B \textbf{71}, 224103 (2005).

\bibitem{Shayegan2003}
    M. Shayegan, K. Karrai, Y. P. Shkolnikov, K. Vakili, E. P. D. Poortere, and S. Manus, Appl. Phys. Lett. \textbf{83}, 5235 (2003).
\bibitem{Botters2006}
    B. Botters, F. Giesen, J. Podbielski, P. Bach, G. Schmidt, L. W. Molenkamp, and D. Grundler, Appl. Phys. Lett. \textbf{89}, 242505 (2006).
\bibitem{Bickford1955}
    L. R. Bickford, J. Pappis, and J. L. Stull, Phys. Rev. \textbf{99}, 1210 (1955).
\bibitem{Lee1955}
    E. W. Lee, Rep. Prog. Phys. \textbf{18}, 184 (1955).
\bibitem{Miao2005}
    G. Miao, G. Xiao, and A. Gupta, Phys. Rev. B \textbf{71}, 094418 (2005).

\bibitem{Gepraegs2010}
    S. Gepr\"{a}gs, A. Brandlmaier, M. Opel, R. Gross, and S. T. B. Goennenwein,
    Appl. Phys. Lett. \textbf{96}, 142509 (2010).
\bibitem{Shebanov1981}
    L. A. Shebanov, phys. stat. sol. (a) \textbf{65}, 321 (1981).
\bibitem{Kay1949}
    H. F. Kay and P. Vousden, Philos. Mag. \textbf{40}, 1019 (1949).
\bibitem{Chopdekar2006}
    R. V. Chopdekar and Y. Suzuki, Appl. Phys. Lett. \textbf{89}, 182506 (2006).
\bibitem{Lee2000}
    M. K. Lee, T. K. Nath, C. B. Eom, M. C. Smoak, and F. Tsui, Appl. Phys. Lett. \textbf{77}, 3547 (2000).
\bibitem{Dale2003}
    D. Dale, A. Fleet, J. D. Brock, and Y. Suzuki, Appl. Phys. Lett. \textbf{82}, 3725 (2003).
\bibitem{Eerenstein2007}
    W. Eerenstein, M. Wiora, J. L. Prieto, J. F. Scott, and N. D. Mathur, Nature Mater. \textbf{6}, 348 (2007).
\bibitem{Tian2008}
    H. F. Tian, T. L. Qu, L. B. Luo, J. J. Yang, S. M. Guo, H. Y. Zhang, Y. G. Zhao, and J. Q. Li, Appl. Phys. Lett. \textbf{92}, 063507 (2008).
\bibitem{Vaz2009}
    C. A. F. Vaz, J. Hoffman, A.-B. Posadas, and C. H. Ahn, Appl. Phys. Lett. \textbf{94}, 022504 (2009).
\bibitem{Serrate2007}
    D. Serrate, J. M. de Teresa, P. A. Algarabel, C. Marquina, J. Blasco, M. R. Ibarra, and J. Galibert, J. Phys.: Condens. Matter \textbf{19}, 436226 (2007).
\bibitem{Kato2002}
    H. Kato, T. Okuda, Y. Okimoto, Y. Tomioka, Y. Takenoya, A. Ohkubo, M. Kawasaki, and Y. Tokura, Appl. Phys. Lett. \textbf{81}, 328 (2002).
\bibitem{Vaitheeswaran2005}
    G. Vaitheeswaran, V. Kanchana, and A. Delin, Appl. Phys. Lett. \textbf{86}, 032513 (2005).
\bibitem{Czeschka2009}
    F. D. Czeschka, S. Gepr\"{a}gs, M. Opel, S. T. B. Goennenwein, and R. Gross, Appl. Phys. Lett. \textbf{95}, 062508 (2009).
\bibitem{Gepraegs2009}
    S. Gepr\"{a}gs, F. D. Czeschka, M. Opel, S. T. B. Goennenwein, W. Yu, W. Mader, and R. Gross, J. Magn. Magn. Mater. \textbf{321}, 2001 (2009).

\end{thebibliography}
%

\end{document}